\documentclass[10pt, conference]{IEEEtran}

\pdfoutput=1 
\usepackage[usenames, dvipsnames]{color}
\usepackage{xcolor}

\usepackage{multirow}
\usepackage{bm}
\usepackage{booktabs}
\usepackage{amsmath}
\usepackage{amsfonts}
\usepackage{multirow}
\usepackage{epsfig}
\usepackage{psfrag}
\usepackage{pifont}
\usepackage{cite}
\usepackage{url}
\usepackage{algorithm}
\usepackage{algorithmic}
\usepackage{graphicx}
\graphicspath{ {./figs/} }
\usepackage[usenames]{color}
\usepackage{amssymb,amsmath}
\usepackage{amsthm}
\usepackage{subcaption}
\usepackage{textcomp}
\usepackage{wrapfig}
\usepackage{floatflt}
\usepackage[utf8]{inputenc}
\usepackage{tabularx}
\usepackage{subcaption}
\usepackage[labelfont=bf,textfont=bf]{caption}

\usepackage{amsmath}
\usepackage{amsfonts}
\usepackage{float}
\usepackage[hidelinks]{hyperref} 
\usepackage{pdfcomment}
\usepackage[usenames, dvipsnames]{color}
\usepackage{xcolor}

\newcounter{numberlistc}

\newcounter{itemlistc}

\newenvironment{itemlist}
    {   \setcounter{itemlistc}{0}
    \begin{list}{$\bullet$}
        {\usecounter{itemlistc}
        \setlength{\parsep}{0pt}
        \setlength{\topsep}{3pt}
        \setlength{\itemsep}{0pt}}
        }{ \end{list} }

\def\BibTeX{{\rm B\kern-.05em{\sc i\kern-.025em b}\kern-.08em
    T\kern-.1667em\lower.7ex\hbox{E}\kern-.125emX}}
 
\small
\usepackage{array}
\newcolumntype{P}[1]{>{\centering\arraybackslash}p{#1}}
\begin{document}

\title{Enhanced Hybrid Temporal Computing Using Deterministic Summations for Ultra-Low-Power Accelerators}


\author{
    \IEEEauthorblockN{Sachin Sachdeva,  Jincong Lu, Wantong Li and Sheldon X.-D. Tan}
    \IEEEauthorblockA{Department of Electrical and Computer Engineering, University of California, Riverside, CA 92521}
    \IEEEauthorblockA{ssach008@ucr.edu, jlu189@ucr.edu, wantong.li@ucr.edu, stan@ece.ucr.edu}
}

\maketitle

\begin{abstract}
This paper presents an accuracy-enhanced Hybrid Temporal Computing (E-HTC) framework for ultra-low-power hardware accelerators using deterministic additions. The framework is inspired by the recently proposed HTC architecture, which leverages pulse-rate and temporal data encoding to reduce switching activity and energy consumption but suffers accuracy loss due to its use of a multiplexer (MUX) for scaled addition. To address this limitation, we propose two bitstream addition schemes: (1) an Exact Multiple-input Binary Accumulator (EMBA), which performs precise binary accumulation over multiple-input bitstreams, and (2) a Deterministic Threshold-based Scaled Adder (DTSA), which employs threshold-based logic for scaled addition. These adders are integrated into a multiplier--accumulator (MAC) unit supporting both unipolar and bipolar encodings. To validate the framework, we implement two hardware accelerators: a Finite Impulse Response (FIR) filter and an 8-point Discrete Cosine Transform (DCT)/iDCT engine.  
Experimental results on a $4\times4$ MAC show that, in unipolar mode, E-HTC matches the RMSE of state-of-the-art Counter-Based Stochastic Computing (CBSC) MAC, improves accuracy by 94\% over MUX-based HTC, and reduces power and area by up to 23\% and 7\% compared to MUX-based HTC and by over 64\% and 74\% compared to CBSC. In bipolar mode, the E-HTC MAC achieves an RMSE of 2.09\%---an 83\% improvement over bipolar MUX-based HTC and approaches bipolar CBSC's RMSE of 1.40\% with area and power savings of up to 28\% and 43\% relative to MUX-based HTC and around 76\% each relative to CBSC. In FIR filter experiments, both E-HTC variants achieve PSNR gains of 3--5\,dB (30--45\% RMSE reduction) over MUX-based HTC, while delivering up to 13\% power and 3\% area savings. Compared to CBSC, E-HTC reduces area by nearly 64\% and power by 62\%. For DCT/iDCT, E-HTC boosts PSNR by 10--13\,dB (70--75\% RMSE reduction) while providing substantial area and power savings over both MUX- and CBSC-based designs.
\end{abstract}

\begin{IEEEkeywords}
EMBA, DTSA, Approximate Computing, Stochastic Computing, MAC Accelerator, Low-Power VLSI
\end{IEEEkeywords}
\section{Introduction}

A significant challenge in modern computing is the exponential increase in computational power driven by emerging artificial intelligence (AI) technologies, contrasted with only linear growth in power supply. This trend suggests that computing energy consumption will roughly double every three years, whereas global energy production increases at a modest linear rate of about 2-3\% per year~\cite{decadal_plan}. Addressing this imbalance requires substantial improvements in energy efficiency.

To achieve drastic reductions in energy consumption, fundamentally new computing paradigms with ultra-low power requirements are needed. One promising direction is temporal computing~\cite{Madhavan:ISCA’14, Madhavan:JETCS’21}, which builds on the concept of race logic~\cite{Madhavan:ISCA’14}. In contrast to conventional binary systems that encode data using voltage levels, race logic encodes information in the timing or delay of signal transitions. Although this approach may introduce some accuracy loss, it offers notable speed, energy efficiency, and area savings benefits. However, its reliance on wavefront-based signal propagation restricts its applicability to spatially structured data, such as trees or graphs, limiting its use in general-purpose computing.


To overcome the limitations of conventional temporal computing, a hybrid temporal computing (HTC) framework was recently introduced~\cite{Maliha:ASPDAC’24}. HTC integrates temporal encoding with bitstream-based (pulse-rate) representations to support ultra-low energy hardware accelerators. Its multiplication scheme accepts two input types: temporal bitstreams (TB), which represent each value by its delay relative to a reference signal, and regulated bitstreams (RB), which distribute ‘1’s deterministically according to their binary weights. This hybrid encoding significantly reduces switching activity, resulting in substantial power and area savings compared to deterministic, stochastic computing approaches such as counter-based stochastic computing (CBSC)~\cite{SimLee:DAC’17, Yu:DAC'21}. Unlike CBSC, HTC supports in-stream bitstream processing across multiple MAC layers, preserving data encoding throughout the computation pipeline. However, the framework incurs higher accuracy loss due to its use of scaled, approximate addition operations using MUX.

In this paper, we mitigate the accuracy issues of HTC and propose two new schemes to perform more accurate additions in the HTC framework while preserving their significant area and power savings.
Specifically, we propose an enhanced HTC (E-HTC) framework over the existing HTC framework. The new E-HTC consists of the following  key contributions summarized below:

\begin{itemlist}


\item We first propose an {\it Exact Multiple-input Binary Accumulator (EMBA)}, which utilizes a binary accumulator for multi-input bitstream addition to achieve exact computation. This approach reformulates addition as a counting operation, conceptually similar to the counter-based stochastic computing (CBSC) method. However, unlike conventional CBSC designs that employ a single counter per multiplier, our design extends counter-based accumulation to support multiple-input bitstreams. This enhancement enables more efficient resource sharing, improving area and power efficiency.

\item Second, we propose a {\it Deterministic Threshold-Based Scaled Adder (DTSA)} to accurately perform the scaled addition of bitstreams. In this design, the addition of the multiple input bitstreams is modeled as an $n$-value divider, where $n$ is the number of input bitstreams. The output bitstream implements floor division by $n$, while the remainder is accumulated and stored in a register for error compensation for accuracy improvement.  This deterministic scheme avoids the need for random number generators and significantly improves accuracy over MUX-based scaled addition. Notably, the DTSA design preserves the in-stream processing capability of the original HTC framework with a significant improvement in accuracy.

\item We present the new multiplier-accumulator architectures (MACs) for the two proposed E-HTC designs, including a specific $4\times4$ MAC implementation utilizing the new adders. The EMBA directly provides the binary output, whereas DTSA  provides output in bitstream, which we can convert into TB format for subsequent HTC computations or in the binary format for single-step HTC computations. So, DTSA preserves the in-stream computation capability of existing MUX-based HTC.

\item Experimental results for a $4\times4$ MAC in unipolar mode shows that both EMBA- and DTSA-based designs achieve an RMSE of 0.52\%, matching CBSC accuracy while delivering a 94\% improvement over the MUX-based HTC MAC. Regarding hardware efficiency, both designs reduce power consumption by over 64\% and area by more than 74\% compared to CBSC. Relative to the MUX-based HTC MAC, the EMBA variant lowers power and area by around 23\% and 7\%, respectively, while the DTSA variant achieves reductions of 5.6\% in power and 6.4\% in area. In bipolar mode, the both EMBA and DTSA MAC designs achieves an RMSE of 2.09\%—an 83\% improvement over bipolar MUX-based HTC and close to bipolar CBSC’s RMSE of 1.40\%—with area and power savings of up to around 28\% and 43\% versus MUX-based HTC and up to approximately 76\% each versus CBSC.

\item We further evaluate the proposed E-HTC framework by implementing two hardware accelerators: a Finite Impulse Response (FIR) filter and an 8-point Discrete Cosine Transform (DCT)/iDCT engine for image and digital signal processing (DSP) applications.
In the FIR filter design, both EMBA- and DTSA-based MAC architectures significantly outperform the MUX-based counterpart by achieving a 3–5\,dB improvement in peak signal-to-noise ratio (PSNR), equivalent to a 30–45\% reduction in RMSE while maintaining similar accuracy to the CBSC-based MAC. Relative to the MUX-based HTC design, the EMBA design reduces area and power by around 3\% and 13\%, respectively, while the DTSA variant achieves 1\% area and 2\% power savings. Compared to the CBSC-based FIR, both proposed E-HTC MAC designs reduce the area by more than 64\%, with power savings of roughly 62\% for EMBA and 57\% for DTSA.

\item For the 8-point DCT filter, the EMBA- and DTSA-based MACs deliver a 10–13\,dB PSNR gain over the MUX-based design, translating to a 70–75\% reduction in RMSE. Although the CBSC-based MAC provides the highest PSNR and lowest RMSE, the E-HTC-based DCT design achieves competitive PSNR while dramatically reducing hardware cost—offering more than 10\% area and nearly 90\% power savings compared to CBSC. Relative to the
MUX-based design, both E-HTC designs achieve more than 8\% area and up to 14\% power reductions.
\end{itemlist}


\section{Preliminary and review of related works}
\label{sec:rela_work}

\subsection{Review of deterministic stochastic computing}
\label{sec:review_of_determ_sc}

\begin{figure}
    \centering
    \includegraphics[width=0.49\textwidth]{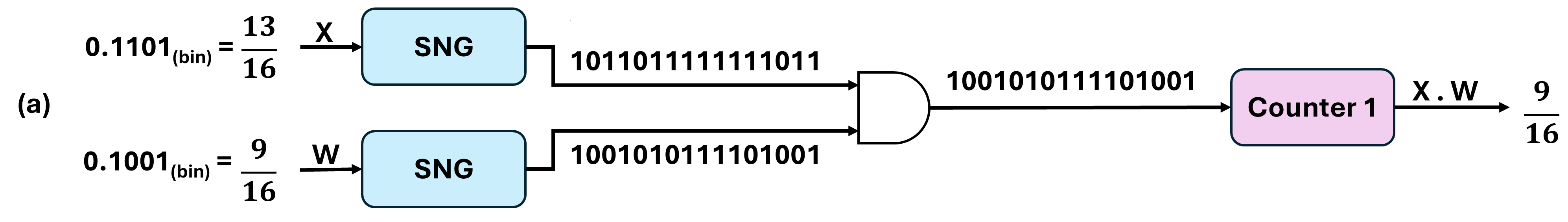}
    
    
    \includegraphics[width=0.49\textwidth]{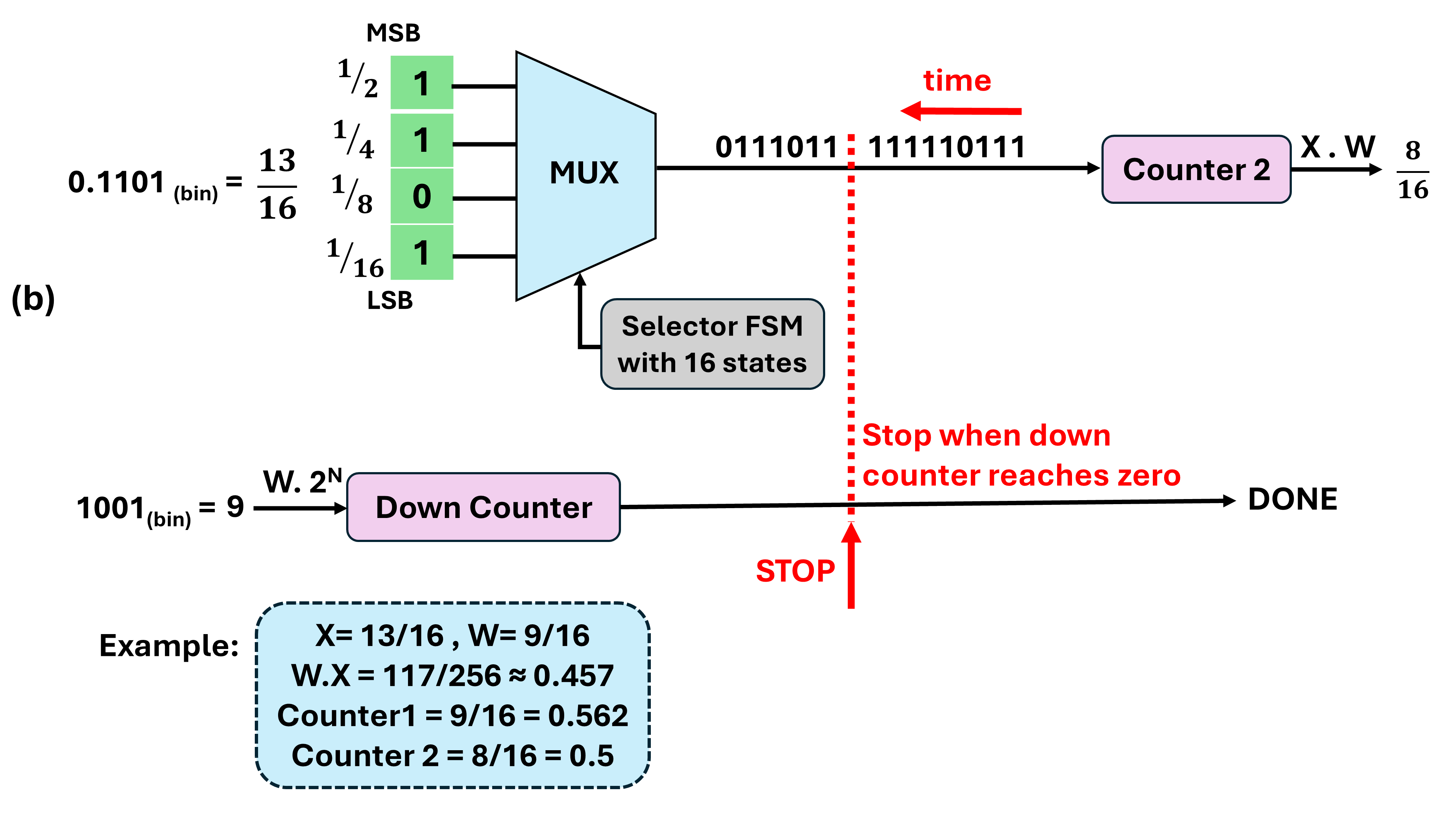}
    
    \caption{(a) Traditional stochastic multiplication; (b) The counting-based stochastic (CBSC) multiplication~\cite{SimLee:DAC’17,Yu:DAC'21}}
    \label{fig:cbsc_mul_concept}
    \vspace{-0.2in}
\end{figure}

Stochastic computing (SC) can be viewed as a specialized form of temporal computing, where binary values are converted into sequential bitstreams using stochastic number generators (SNGs). These generators rely on random number sources to compare against fixed binary inputs, producing probabilistic bitstreams~\cite{Gaines:Book'69, Qian:journal'2010, Alaghi:TECS'13}. A key advantage of SC is its ability to implement complex arithmetic using simple logic. For instance, multiplication in unipolar coding can be realized with a basic {\it AND} gate, as illustrated in Fig.\ref{fig:cbsc_mul_concept}(a). However, this simplicity comes at the cost of accuracy, as SC is vulnerable to random fluctuations and bitstream correlation effects\cite{Alaghi:ICCD'13}.

Recent advancements in counting-based stochastic computing (CBSC) have eliminated the need for random bitstreams in multiplication without sacrificing accuracy~\cite{SimLee:DAC’17, Yu:DAC'21}. As shown in Fig.~\ref{fig:cbsc_mul_concept}(b), CBSC replaces the conventional stochastic {\it AND}-based multiplication with a deterministic counting mechanism. In this approach, we count exactly $\lvert W\rvert$ bits of input~$X$. This idea uses a down counter controlled by input $W$.

Unlike traditional stochastic multiplication, CBSC requires only a single finite state machine (FSM)-based stochastic number generator (SNG) to convert one binary input—typically $X$—into a deterministic bitstream. The FSM-based SNG ensures an even distribution of each bit $X_{i-1}$ according to its binary weight $2^{i-1}$. For example, if $i = 4$, the bit $X_3$ appears 8 times in the generated stochastic number. We implement this structured bitstream efficiently using an FSM and a multiplexer (MUX), as illustrated in Fig.~\ref{fig:cbsc_mul_concept}(b). The FSM operates as an up-counter from 0 to $2^N-1$ (for an $N$-bit input $X$). At the same time, the MUX selects the appropriate $X_{i-1}$ bit based on the current counter value, generating a deterministic and uniformly distributed output bitstream.


For the second input, $W$, if we reorder the bits such that all the  ‘1’s appear first, it does affect the value of SN for $W$ or that of resulting SN after the AND operation~\cite{SimLee:DAC’17}. Since SC multiplication relies solely on the {\it AND} operation, only the initial portion of the output bitstream—corresponding to the ‘1’s in $W$—needs to be processed. As a result, the counting process requires only $W \cdot 2^N$ cycles, effectively reducing the average latency by half. To implement this, the authors in~\cite{SimLee:DAC’17} introduced a down counter initialized to $W \cdot 2^N$, which decrements by one each clock cycle until it reaches zero, signaling the end of the computation. As a result, CBSC leads to a simpler design, as one traditional SNG, typically implemented using linear feedback shift registers (LFSR) along with the AND gate are removed in exchange for a down counter, which is much cheaper than an SNG. The complete CBSC multiplication design is illustrated in Fig.~\ref{fig:cbsc_mul_concept}(b).

CBSC has evolved beyond its stochastic origins and is now better understood as a specialized form of temporal computing, no longer dependent on stochastic processes. It forms the conceptual foundation for the recently proposed hybrid temporal computing (HTC) paradigm~\cite{Maliha:ASPDAC’24}. Subsequent research has enhanced CBSC~\cite{Chen:ICCAD'20,Yu:DAC'21,Yu:DAC'22}, including improvements in the counting mechanism~\cite{Yu:DAC'21}, the development of scaled implementations, and the design of high-accuracy scaled CBSC multipliers for improved numerical accuracy~\cite{Yu:DAC'22}. However, CBSC still needs to be extended to complete arithmetic units—such as adders, multiplier–accumulators (MACs), or general processing elements—primarily because the counting process typically operates in binary format, a characteristic originally referred to as binary-interface stochastic computing~\cite{SimLee:DAC’17}.

    
    
    
Beyond CBSC, several deterministic stochastic computing (SC) designs have been proposed in recent years. Early methods replicated one operand’s bitstream for each bit of the other operand~\cite{Jenson:ICCAD'2016, Najafi:TVLSI'2019}, resulting in quadratic growth in bitstream length. More recent approaches address this scalability issue using bitstream length reduction and approximation techniques to reduce overhead~\cite{Kiran:GLSVLSI'22}.

\subsection{Review of hybrid temporal computing (HTC)}
\label{sec:review_of_htc}

The Hybrid Temporal Computing (HTC) framework was recently proposed~\cite{Maliha:ASPDAC’24}, building upon the temporal data encoding principles of race logic~\cite{Madhavan:ISCA’14, Madhavan:JETCS’21}. In race logic, information is encoded through the relative timing of rising edges across different wires, referencing a common temporal baseline. This scheme enables each wire to convey multiple bits of information based on the arrival time of a single transition, resulting in low switching activity and high energy efficiency. However, pure temporal computing remains limited in its ability to support general arithmetic operations—such as multiplication and addition—due to constraints imposed by its waveform-based data representation.
 
\begin{figure}[ht!]
     \includegraphics[width=0.49\textwidth]{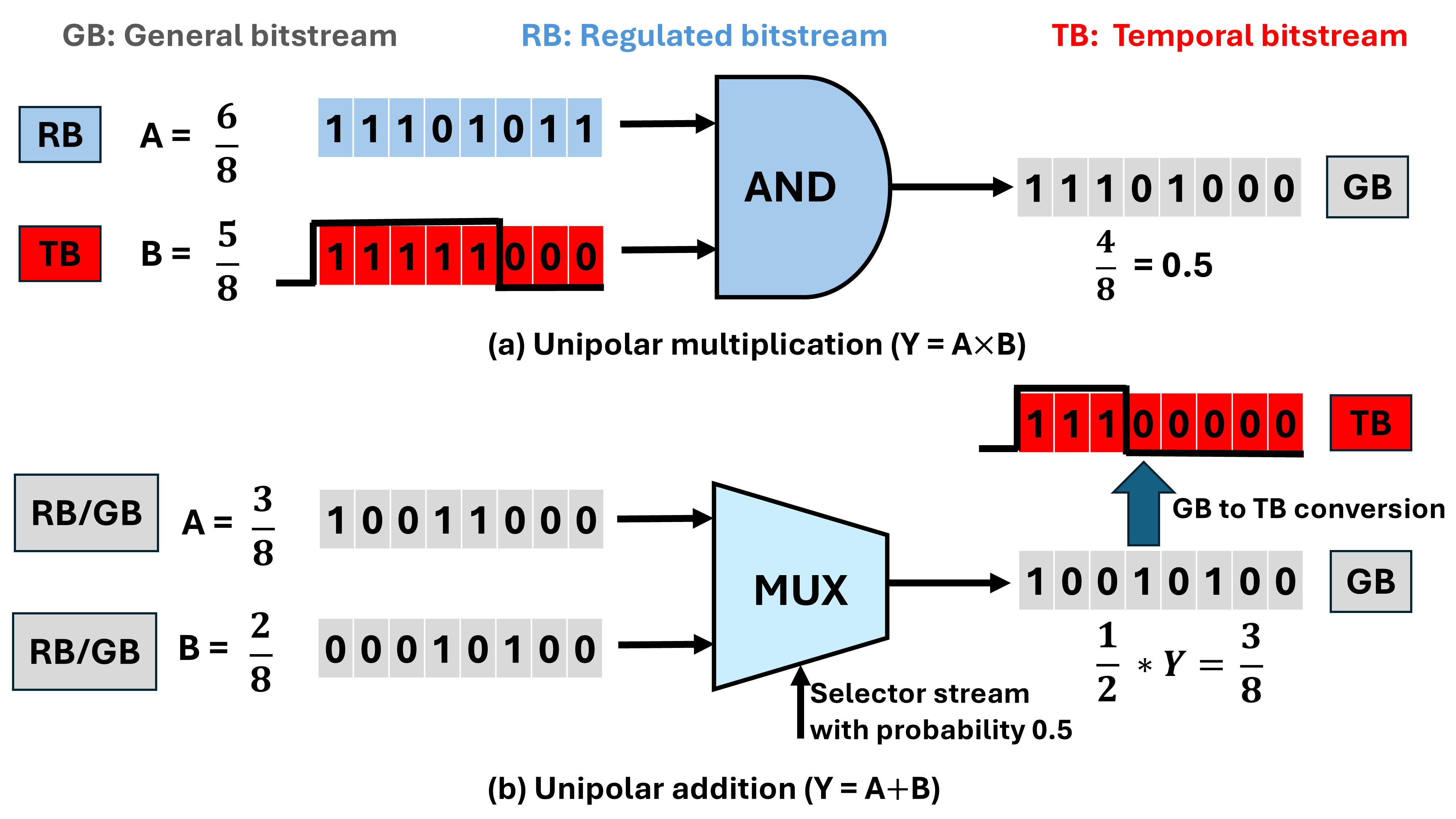}
      \includegraphics[width=0.49\textwidth]{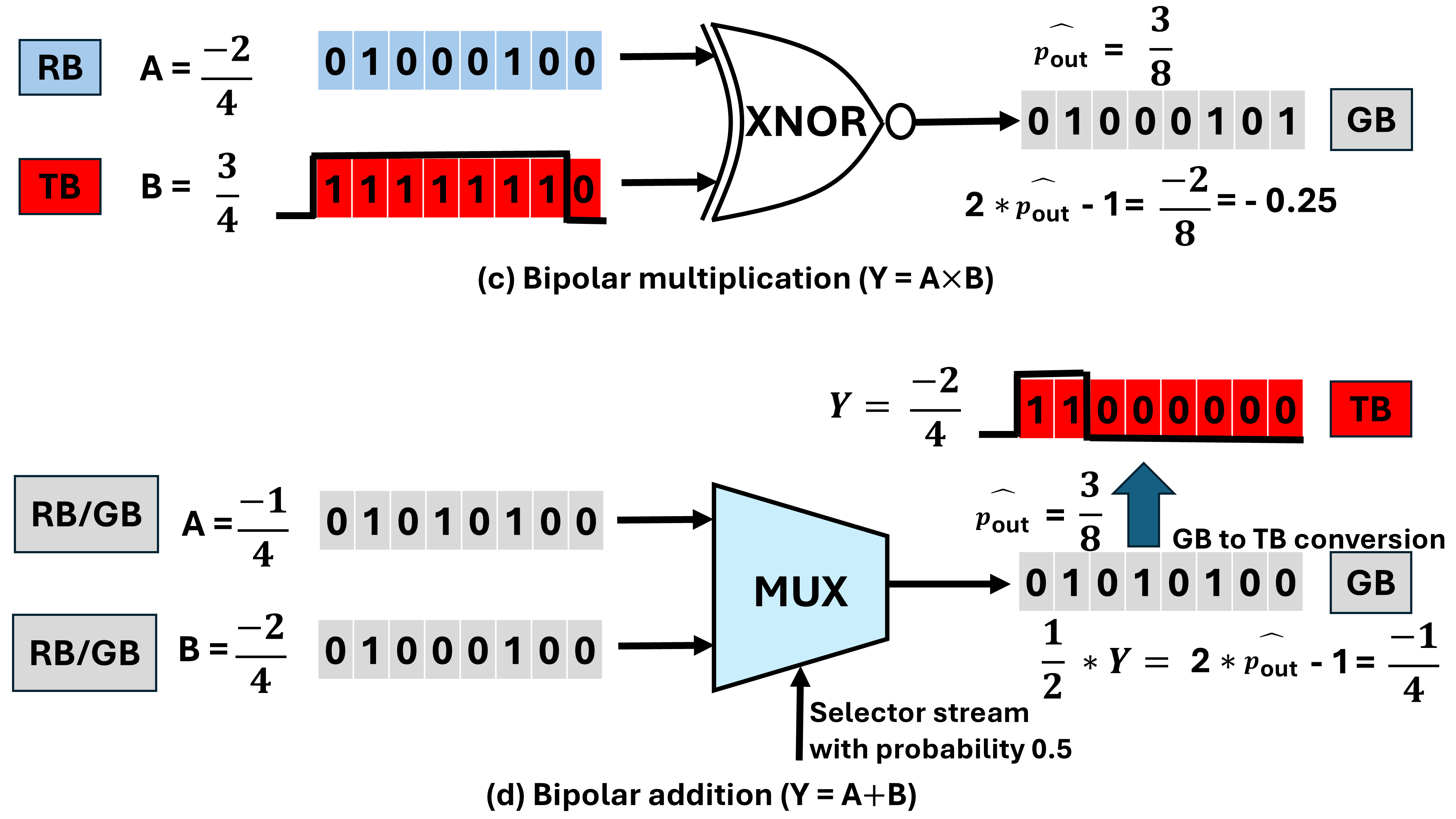}
 \caption{The multiplication and addition in the HTC framework}
  \label{fig:htc_multiply_and_add} 
   \vspace{-0.2in}
\end{figure}

To overcome the limitations of pure temporal computing, the Hybrid Temporal Computing (HTC) framework was proposed in~\cite{Maliha:ASPDAC’24}. HTC encodes data using three distinct bitstream formats, as illustrated in Fig.~\ref{fig:htc_multiply_and_add}, each with uniform bit weights. The formats include: (1) the General Bitstream (GB) format, which corresponds to the traditional stochastic bitstream; (2) the Regulated Bitstream (RB) format, where the ‘1’s are evenly distributed across the bitstream based on their binary weight; and (3) the Temporal Bitstream (TB) format, where the value is represented by a time delay relative to a reference signal. These encoding schemes support unipolar representation (values in $[0,1]$) and bipolar representation (values in $[-1,1]$), offering flexibility for a wide range of arithmetic and signal processing applications.

In the RB bitstream representation, the '1's are distributed across the bitstream according to its binary weight with a length of $2^N$ for a binary number of \(N\). For unipolar values in \([0,1]\), consider the 3‑bit binary number \(110_2\) (decimal 6) is encoded in an 8-bit RB as 11101011 which represents \(6/8\). Conversely, unipolar temporal bitstream (TB) encoding packs all ‘1’s at the front of the stream, followed by the remaining ‘0’s, representing bitstream as 11111100. Bipolar RB encoding extends the same concept to signed values in \([-1,1]\) by first mapping the unipolar fraction by calculating the required number of ‘1’s as \(p = \frac{X + 1}{2}\). As an illustration, consider a 3-bit signed binary number \(110_2\), which corresponds to -2 or scaled number $X=$\(-2/4\) which yields \( p = \frac{(-2/4) + 1}{2} = 2/8 \), requiring two ‘1’s in an 8‑bit bitstream. Distributing these two ‘1’s according to the RB rule might produce the sequence as 01000100. Bipolar TB, on the other hand, again groups all ones at the head of the stream, yielding 11000000.
  
In the HTC multiplication, one input is provided in the RB format and the other in the TB format to an {\it AND/XNOR} gate for unipolar/bipolar encoding. In unipolar, we count the fraction of ‘1’s in the output bitstream (\(\hat p_{\mathrm{out}} = \frac{\#\text{ones}}{N}\)) to get the output. However, In the bipolar scenario, we count the fraction of ‘1’s, \(\hat p_{\mathrm{out}}\), as before, then we recover the bipolar output as \(2\hat p_{\mathrm{out}} - 1\). This HTC configuration reduces input switching activity, resulting in lower energy consumption than traditional stochastic computing. In addition, HTC employs a scaled adder, implemented using a multiplexer (MUX) controlled by a random signal with a 0.5 probability of ‘1’—suitable for two-input averaging. However, this approach introduces approximation errors and incurs additional area overhead due to the need for random number generation, making it less accurate than the CBSC method. After arithmetic operations, we can transform the output bitstream into the TB format for energy-efficient data propagation to subsequent stages. The idea of HTC multiplication and addition in unipolar and bipolar encoding is presented with examples in Fig.~\ref{fig:htc_multiply_and_add}.

As noted earlier, MUX-based scaled approximate addition incurs accuracy loss, motivating exploration of deterministic accumulation primitives in prior work. Accumulative parallel counters (APC)~\cite{Parhami:ACSSC'95} provide a generic mechanism for summing multiple 1-bit inputs per cycle and output the corresponding binary number accumulating the result, and have recently been adopted in SC-DCNN~\cite{Ao:ACM'17} using approximate variants to reduce hardware cost at the expense of small accuracy loss. In contrast, our Exact Multiple-input Binary Accumulator (EMBA) retains exact counting but specializes the primitive for HTC’s RB\&TB streams by deriving HTC encoding-aware adder and accumulator sizing as a function of fan-in $M$ and bitstream length $N$, ensuring overflow-free and bias-free accumulation. Our proposed design will be described in detail in Section~\ref{sec:architecture}.

In a similar direction, uGEMM~\cite{Di:ISCA'20} proposed the unary scaled adder (uSADD) to realize deterministic scaled addition in unary computing. uSADD accumulates inputs against a fixed threshold but discards the residual, guaranteeing only bounded error in the output stream. By contrast, our DTSA retains and reinserts the residual during binary readout, enabling exact reconstruction of the MAC result while still supporting in-stream operation within HTC pipelines. The full design will be discussed in the Section~\ref{sec:architecture}.

\vspace{0.13in}
\section{The enhanced hybrid temporal computing (E-HTC) framework}
\label{sec:architecture} 
\vspace{-0.04in}
In this section, we present the fundamental arithmetic computing units for the enhanced HTC (E-HTC) framework. These include multipliers, accumulators, and multiplier-accumulators (MACs). 
\vspace{-0.04in} 
\subsection{New deterministic addition in E-HTC}
In the {\it Enhanced Hybrid Temporal Computing (E-HTC)} framework, data continues to be represented using three distinct formats, as shown in Fig.~\ref{fig:htc_multiply_and_add}. Multiplication follows the same strategy as in the original HTC approach\cite{Maliha:ASPDAC’24}, where inputs in regulated bitstream (RB) and temporal bitstream (TB) formats are processed through an {\it AND/XNOR} gate for unipolar/bipolar encoding. The key advancement in E-HTC lies in introducing new, deterministic adder designs. In the following subsections, we present two such adders that significantly improve accuracy and hardware efficiency.

\subsubsection{Exact Multiple-Input Binary Accumulator (EMBA)}
\label{subsec:epau}

\begin{figure}[htbp!]
  \centering
  \includegraphics[width=.95\columnwidth]{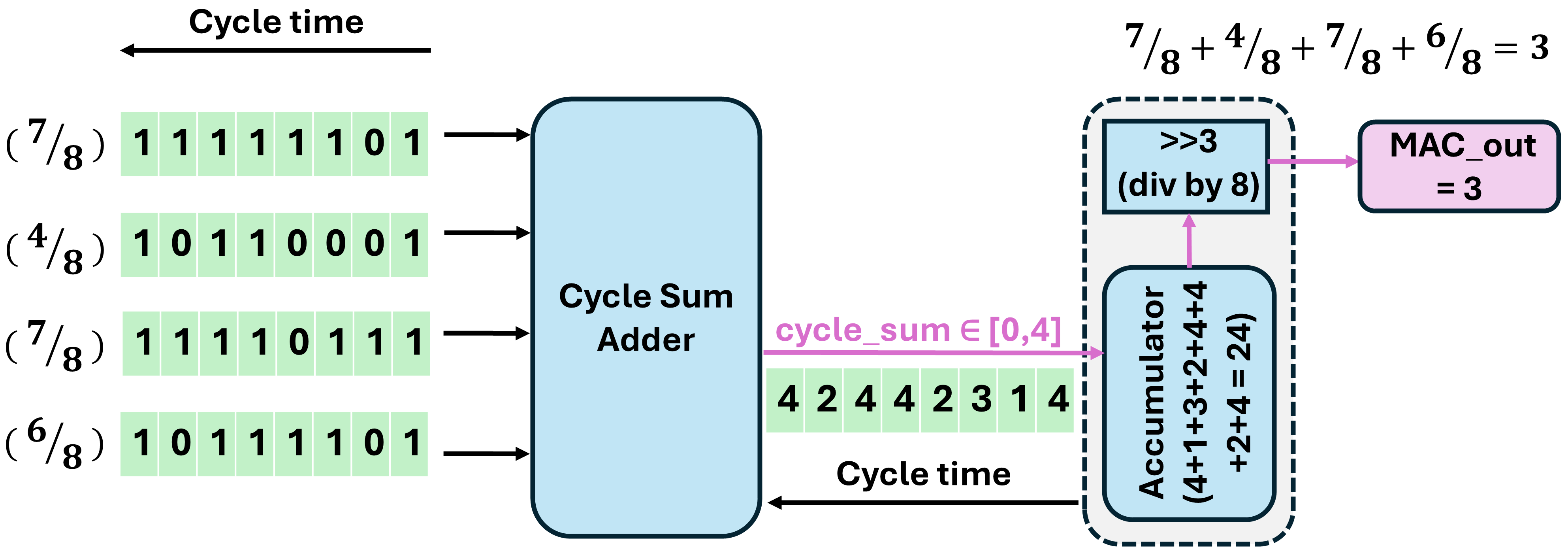}
  \caption{Block diagram of the Exact Multiple‑Input Binary Accumulator (EMBA). The Cycle Sum Adder computes the sum of four 1‑bit product outputs per cycle, which is accumulated over 8 cycles. The final output is derived by dividing the accumulated sum by 8, yielding the MAC output in the range~\(\bm{[0,4]}\).}
  \label{fig:epau_adder}
\end{figure}

The first idea is to use a binary adder, which is exact. In the HTC computing paradigm, each multiplier produces a 1-bit product output every clock cycle. For example, when operating at 3-bit binary precision, the bit stream length is \(2^3 = 8\) cycles. In an 8-input MAC, four multipliers yield four bitstreams (one per multiplier) of length 8, representing the partial products that need to be accumulated. Fig.~\ref{fig:epau_adder} illustrates the proposed EMBA logic. 

Let the four product output bits at cycle \(c \in \{0,\dots,7\}\) be denoted by \(\bigl(M_1(c),\,M_2(c),\,M_3(c),\,M_4(c)\bigr)\),
where \(M_i(c) \in \{0,1\}\). In contrast to existing MUX-based scaled addition, which samples a single product bit from each cycle, the EMBA approach deterministically aggregates all four bits at every cycle using a parallel \emph{Cycle Sum Adder}. Specifically, we define the cycle sum as
{\setlength{\abovedisplayskip}{2pt}%
 \setlength{\belowdisplayskip}{2pt}%
 \begin{equation}
   \text{cycle\_sum}(c) \;=\; \sum_{i=1}^{4} M_i(c), \quad c=0,\dots,7
   \label{eq:epau_cycle_sum}
 \end{equation}%
}
where \(\text{cycle\_sum}(c) \in \{0,1,2,3,4\}\). At each cycle, the Cycle Sum Adder adds \(\text{cycle\_sum}(c)\) to a running \emph{accumulator} register:
{\setlength{\abovedisplayskip}{2pt}%
 \setlength{\belowdisplayskip}{3pt}%
 \begin{equation}
   \text{accumulator}
     \;=\;
   \text{accumulator}
     \;+\;
   \text{cycle\_sum}(c)
   \label{eq:epau_accum}
 \end{equation}%
}
After eight cycles, the accumulator holds the total count of high product bits (‘1’s). Since each cycle can contribute at most 4, the maximum possible accumulator value is \(4\times8=32\), which requires a
$\lceil \log_{2}(32+1) \rceil = 6$ bit
register. Finally, under unipolar stochastic encoding (with values in \([0,1]\)), the final MAC output is obtained by scaling the accumulator value:
\begingroup
\setlength{\abovedisplayskip}{4pt}%
\setlength{\belowdisplayskip}{4pt}%
\begin{equation}
\text{MAC}_{\text{out}} \;=\; \frac{\text{accumulator}}{8} \quad \in \; [0,4].
\label{eq:epau_mapping}
\end{equation}
\endgroup
In hardware, the divide-by-8 operation in binary is efficiently implemented by a right shift operation of 3 bits.

Although EMBA is illustrated with a four-input example, the architecture scales naturally to larger fan-in by either increasing the adder and accumulator sizes or by tiling multiple EMBA blocks and hierarchically combining their results. Specifically,  for $M$ inputs and a stream length of $N$, the adder must be configured with $\lceil \log_{2}(M+1) \rceil$ bits to represent the maximum per-cycle sum of $M$, and the accumulator must be configured with $\lceil \log_{2}(MN+1) \rceil$ bits to represent the maximum accumulated value over $N$ cycles.

By deterministically aggregating all four product bits in each cycle, the EMBA approach eliminates the random-induced accuracy loss inherent in MUX-based adders. Consequently, it achieves significantly higher accuracy while also reducing power consumption and area due to its streamlined design. In the result section, we will demonstrate the enhanced accuracy and hardware efficiency of E-HTC MAC designs when integrated with the proposed EMBA logic.

\subsubsection{Deterministic Threshold-Based Scaled Adder (DTSA)}
\label{subsec:dtsa}
The second approach (DTSA) is to perform scaled addition by accumulating all four product bits in each clock cycle. The idea of DTSA has been described in Fig.~\ref{fig:dtsa_adder}.
\begingroup
  \setlength{\abovecaptionskip}{2pt}
  \setlength{\belowcaptionskip}{2pt}
\begin{figure}[h!]
  \centering
  \includegraphics[width=.95\columnwidth]{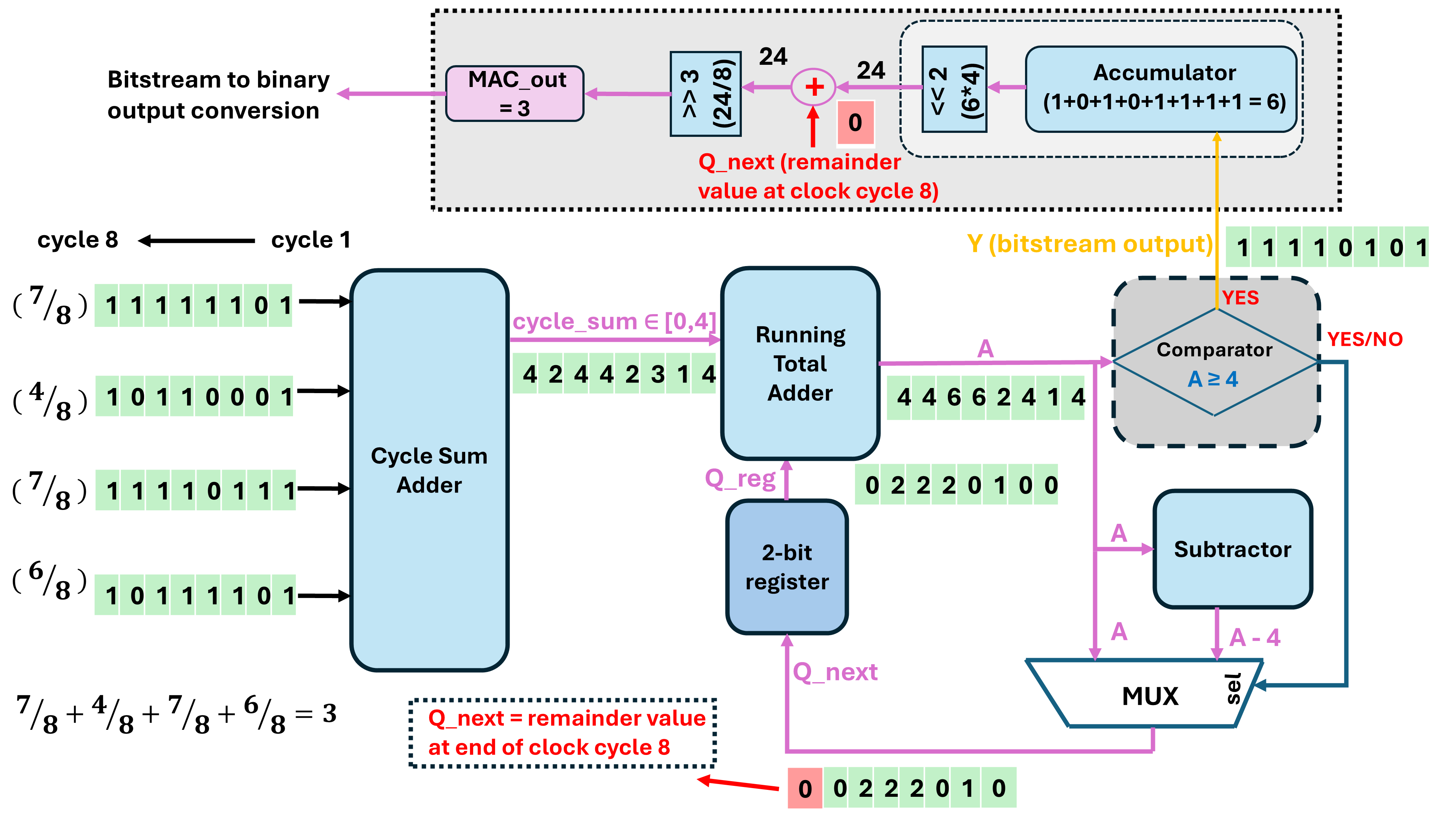}
  \caption{Block diagram of the Deterministic Threshold-based Scaled Adder (DTSA). In each cycle, the four 1-bit product outputs are summed and added to a 2-bit accumulator. When the sum reaches or exceeds 4, a ‘1’ is generated, and 4 is subtracted, effectively implementing a floor operation over 8 cycles. The final MAC output in the range~\(\bm{[0,4]}\) is obtained by combining the accumulated count with any residual and dividing the total by 8.}
  \label{fig:dtsa_adder}
  \vspace{-0.1in}
\end{figure}
{\setlength{\textfloatsep}{1pt}
 \captionsetup[table]{aboveskip=2pt, belowskip=2pt}

 \renewcommand{\arraystretch}{1.3}
 \setlength{\tabcolsep}{8pt}
\vspace{-6pt}
 \begin{table}[htbp!]
   \centering
   \scriptsize
   \resizebox{\columnwidth}{!}{%
     \begin{tabular}{*{10}{c}}
       \toprule
       \multicolumn{1}{c}{\textbf{Cycle}}
       & \multicolumn{1}{c}{$\mathbf{M}_1$}
       & \multicolumn{1}{c}{$\mathbf{M}_2$}
       & \multicolumn{1}{c}{$\mathbf{M}_3$}
       & \multicolumn{1}{c}{$\mathbf{M}_4$}
       & \multicolumn{1}{c}{\textbf{Cycle Sum}}
       & \multicolumn{1}{c}{$\mathbf{Q}_{\mathrm{reg}}$}
       & \multicolumn{1}{c}{$\mathbf{A} = \mathbf{Q}_{\mathrm{reg}} + \textbf{Cycle Sum}$}
       & \multicolumn{1}{c}{$\mathbf{Y}$}
       & \multicolumn{1}{c}{$\mathbf{Q}_{\mathrm{next}}$} \\
       \midrule
       1 & 1 & 1 & 1 & 1 & 4 & 0 (initial) & 4 & 1 & 0 \\
       2 & 0 & 0 & 1 & 0 & 1 & 0 & 1 & 0 & 1 \\
       3 & 1 & 0 & 1 & 1 & 3 & 1 & 4 & 1 & 0 \\
       4 & 1 & 0 & 0 & 1 & 2 & 0 & 2 & 0 & 2 \\
       5 & 1 & 1 & 1 & 1 & 4 & 2 & 6 & 1 & 2 \\
       6 & 1 & 1 & 1 & 1 & 4 & 2 & 6 & 1 & 2 \\
       7 & 1 & 0 & 1 & 0 & 2 & 2 & 4 & 1 & 0 \\
       8 & 1 & 1 & 1 & 1 & 4 & 0 & 4 & 1 & 0 (remainder) \\
       \bottomrule
     \end{tabular}%
   }
   \caption{DTSA Cycle-by-Cycle Computation Example}
   \vspace{-6pt} 
   \label{tab:DTSA_example}
 \end{table}
}
The cycle-by-cycle computation for the same example has been described in Table~\ref{tab:DTSA_example}. As shown in Fig.~\ref{fig:dtsa_adder}, four 1-bit outputs from the multipliers are summed by a \emph{Cycle Sum Adder} (a 3-bit adder that produces values in \(\{0,1,2,3,4\}\)). Specifically, let define $\text{cycle\_sum}(c)$ as
\begingroup
\setlength{\abovedisplayskip}{1pt}%
\setlength{\belowdisplayskip}{1pt}%
\begin{equation}
\text{cycle\_sum}(c) \;=\; \sum_{i=1}^{4} M_i(c), \quad c=0,\dots,7
\label{eq:epau_cycle_sum}
\end{equation}
\endgroup
where each \(M_i(c)\) is a product bit from the \(i\)-th multiplier at cycle \(c\). This sum is added to a 2-bit accumulator register \(Q_{\text{reg}}\) storing the residual \(\{0,1,2,3\}\) for later error compensation. Formally,
\begingroup
\setlength{\abovedisplayskip}{1pt}%
\setlength{\belowdisplayskip}{1pt}%
\begin{equation}
A \;=\; Q_{\text{reg}} \;+\; \text{cycle\_sum}(c).
\label{eq:your_label}
\end{equation}
\endgroup
If \(A \ge 4\), the DTSA subtracts 4 from \(A\) and emits a `1' in that cycle’s output stream \(Y\); otherwise, it emits `0' and retains \(A\). Mathematically,
\begingroup
\setlength{\abovedisplayskip}{2pt}%
\setlength{\belowdisplayskip}{2pt}%
\[
Q_{\text{next}} \;=\;
  \begin{cases}
    A - 4, & \text{if } A \ge 4,\\
    A,     & \text{otherwise,}
  \end{cases}
  \quad
Y[i] \;=\;
  \begin{cases}
    1, & \text{if } A \ge 4,\\
    0, & \text{otherwise.}
  \end{cases}
\]
\endgroup
As we can see, \(Q_{\text{reg}}\) counts all the '1's, which has not produced any '1' in the output $Y$ in the current clock cycle. But it will generate '1' as output when the accumulated value reach or become larger than 4 in the future clock cycles.

Over 8 cycles, each multiplier \(M_k\) generates 8 single-bit products \(M_k[i]\). The total product count is thus
\begingroup
\setlength{\abovedisplayskip}{1pt}%
\setlength{\belowdisplayskip}{1pt}%
\[
\text{Total Product Count} = \sum_{i=0}^{7} \Bigl( M_1[i] + M_2[i] + M_3[i] + M_4[i] \Bigr)
\]
\endgroup
Since each threshold operation effectively divides the total product count by 4, the number of ones in the output bit stream \(Y\) is
\begingroup
\setlength{\abovedisplayskip}{1pt}%
\setlength{\belowdisplayskip}{1pt}%
\begin{equation}
\#Y \;=\; \left\lfloor \frac{\text{Total Product Count}}{4} \right\rfloor.
\end{equation}
\endgroup


with remainder after processing 8 cycles is stored in the accumulator register \(Q_{\text{next}}\), thus implementing floor division.
Finally, to recover the binary MAC output, we add back the remainder in \(Q_{\text{next}}\) by computing:
\begingroup
\setlength{\abovedisplayskip}{2pt}%
\setlength{\belowdisplayskip}{2pt}%
\[
\text{Final\_\#ones}
  \;=\; (\#Y \times 4) \;+\; Q_{\text{next}},
\]
\endgroup
Here, the multiplication by 4 operation is performed with left shift by 2 operation in the hardware as described in the Fig. ~\ref{fig:dtsa_adder}. Finally, we divide this by 8 to get binary MAC output as
\begin{equation}
\text{MAC}_{\text{out}} \;=\; \frac{\text{Final\_\#ones}}{8} \quad \in \; [0,4].
\label{eq:epau_mapping}
\end{equation}
By eliminating the random MUX+LFSR from the accumulator stage, the DTSA provides accurate scaled accumulation via a deterministic flooring operation along with remainder value (compensation term) in the register,  significantly improving precision while maintaining a simple, low-power design.

Again, DTSA is illustrated with a 4-input, 8-cycle case for clarity, the architecture generalizes to arbitrary input sizes and stream lengths. For $M$ inputs, the Cycle-Sum adder must be configured with $\lceil \log_{2}(M+1) \rceil$ bits to represent the maximum per-cycle sum, the residual register requires $\lceil \log_{2}(M) \rceil$ bits to retain the bounded remainder, and the output accumulator only needs $\lceil \log_{2}(N+1) \rceil$ bits since it aggregates the emitted 0/1 stream over $N$ cycles.

\section{Implementation of multiple-input E-HTC MAC design}
\label{proposed_idea_sec}

This section presents detailed hardware implementation for multiplication and accumulation (MAC) operation. MAC operation essentially is to perform $a \leftarrow  a + (b \times c)$ operation. One may have multiple multiplications and one multi-input summation (also called dot product) as follows:
\begingroup
\setlength{\abovedisplayskip}{1pt}%
\setlength{\belowdisplayskip}{1pt}%
\begin{equation}
    a =  \sum_{i=1}^{N} ( b_i \times  c_i)
    \label{eq:mul_mac_op}
\end{equation}
\endgroup
Fig.~\ref{fig:htc_dot_product} shows the general architecture for the multiple-input MAC operations in Eq.~\ref{eq:mul_mac_op},   which serves as a fundamental computing block for DSP and AI computing. In the E-HTC framework, each multiplier takes one input in RB format and another in TB format. Afterward, the E-HTC adder (EMBA or DTSA) performs the addition operation for multiple bitstreams. The EMBA directly provides the binary output, whereas DTSA provides bitstream as output, which can either be converted into TB format for subsequent HTC computations or directly converted into binary.

\begin{figure}[htbp!]
    \centering
    \includegraphics[width=0.4\textwidth]{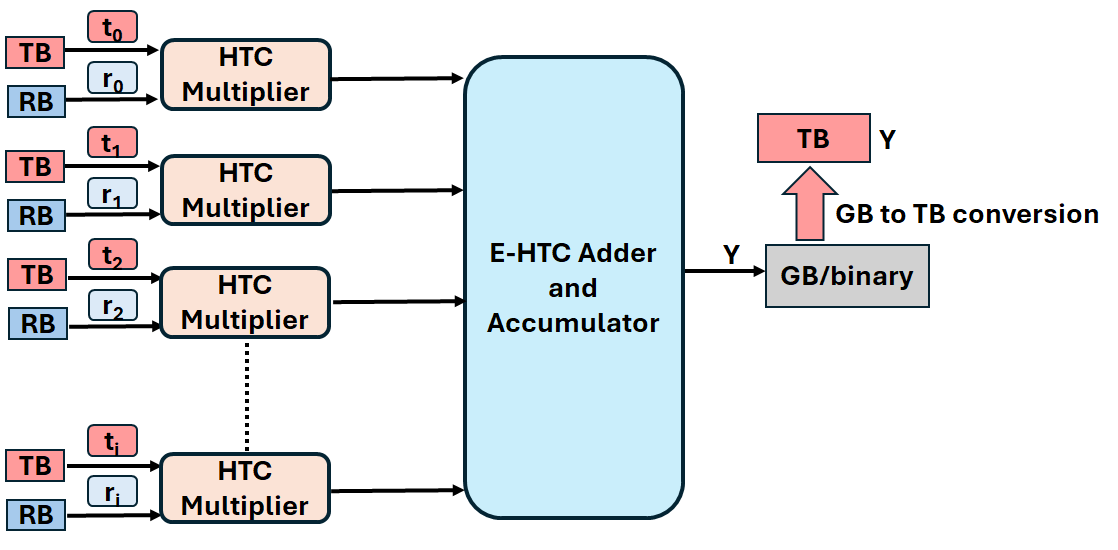}
 \caption{General architecture for the proposed multiple-input E-HTC MAC}
 \label{fig:htc_dot_product}
\end{figure}

In the following, we present a specific MAC design with \( N = 4 \) in Eq.~\ref{eq:mul_mac_op}. The core functional components of the $4\times4$ MAC unit, as illustrated in Fig.~\ref{fig:new_threshold_logic},~\ref{fig:emba_htc_mac}, consist of four HTC multipliers and the proposed E-HTC addition units (EMBA or DTSA). Each HTC multiplier comprises an {\it AND/XNOR} gate for unipolar/bipolar multiplication. This multiplier gate processes one input bitstream from a temporal bitstream (TB) generator ($\,t_i\,$) and another input bitstream from a regulated bitstream (RB) generator ($\,r_i\,$). The RB and the TB of all multiplicands ($X_{b_i}/Y_{b_i}$ in the binary domain) are generated with a single up-counter. 

The output bitstream from each HTC multiplier is then fed into an adder unit (EMBA or DTSA) to perform accumulation. In the EMBA-based HTC MAC, a multi-input binary adder accumulates the bitstreams and produces a binary output. In contrast, the DTSA-based HTC MAC allows flexibility in post-processing: we can either regenerate the binary value or convert the output bitstream into a temporal bitstream (TB) for downstream E-HTC stages.
\subsection{HTC MAC with Deterministic Threshold-Based Scaled Addition}
\label{sec:HTC_MAC_threshold}
The detailed architecture of $4\times4$ E-HTC MAC using DTSA is described in Fig.~\ref{fig:new_threshold_logic}. As we can observe, each HTC multiplier takes one input in RB and another in TB format. These RB and TB bitstreams are generated using a single up-counter for all the multiplicands. The output bitstream of each HTC multiplier is directed to the DTSA adder, which performs the scaled addition of the 4 product output bitstreams using threshold-based logic. 
\begin{figure}[htbp!] \centering
  \includegraphics[width=1\columnwidth]{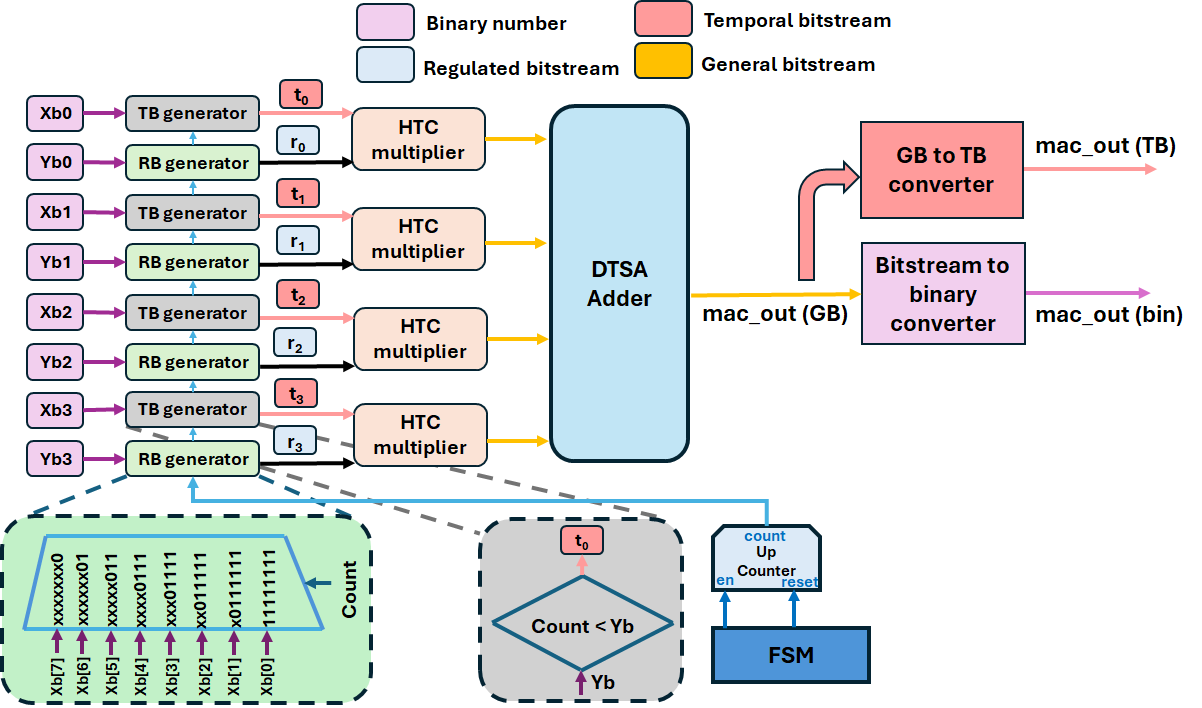}
  \caption{Proposed 4$\times$4 DTSA E-HTC MAC design}
  \label{fig:new_threshold_logic}
\end{figure}
The operating principle of the DTSA is described in detail in the Subsection~\ref{subsec:dtsa}. Notably, the DTSA can produce a binary output that can be used directly in a one-stage MAC computation (via the bitstream-to-binary conversion logic shown in Fig.~\ref{fig:dtsa_adder}), or we can convert the output bitstreams (GB) into the TB format using straightforward shift registers for subsequent MAC computation stages.

\subsection{HTC MAC with exact multiple-input binary accumulator (EMBA)}
\label{sec:HTC_MAC_exact}

In contrast to the DTSA design, which employs per-cycle threshold-based logic for the scaled addition, our second approach—the \emph{Exact Multiple Input Binary Accumulator (EMBA)}-based MAC—adopts a different strategy. The complete $4\times4$ EMBA-based E-HTC MAC architecture is illustrated in Fig.~\ref{fig:emba_htc_mac}. Aside from replacing the adder unit with the EMBA adder, the rest of the architecture remains similar to DTSA-based E-HTC MAC. The detailed architecture of the EMBA adder is depicted in Fig.~\ref{fig:epau_adder}, which utilizes an exact binary adder to perform per-cycle binary additions, and the resulting cycle sums are accumulated in a register. This process precisely counts the number of ones in the input bitstreams, yielding an accurate binary output after one stage of MAC computation. 

\begin{figure}[h!] \centering
  \includegraphics[width=1\columnwidth]{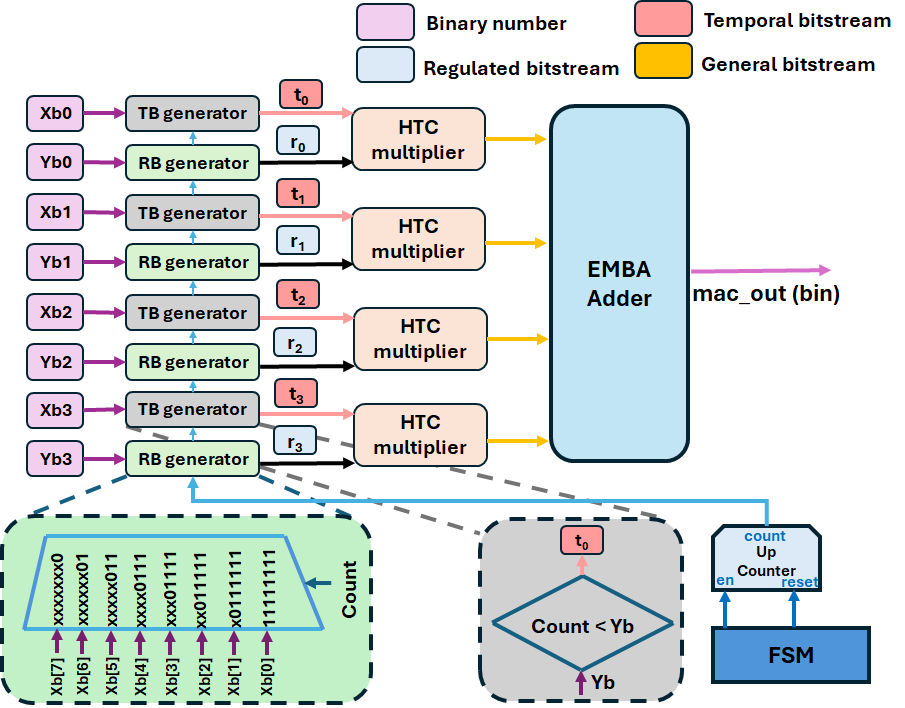}
  \caption{Proposed 4$\times$4 EMBA HTC MAC design}
  \label{fig:emba_htc_mac}
  \vspace{-0.2in} 
\end{figure}

\section{Experimental Results and Discussion}
\label{sec:results}

We benchmark the proposed EMBA‐ and DTSA‐based E-HTC MACs against two state-of-the-art designs: (i) the CBSC-based MAC because CBSC is widely regarded as one of the most accurate deterministic stochastic-computing approaches, with proven area and power advantages over conventional binary MACs—and the (ii) MUX-based HTC MAC~\cite{Maliha:ASPDAC’24}, its direct predecessor in the hybrid-temporal paradigm. 


All four $4\times4$ MAC variants—EMBA, DTSA, CBSC, and MUX-HTC—are first benchmarked as standalone dot-product engines. At the MAC level, both the unipolar and bipolar variants complete a single 256-cycle stochastic pass, corresponding to 2.56 µs at a 10ns clock period. These architectures can scale to larger arrays (e.g., $8\times8$, $16\times16$) by tiling identical $4\times4$ MAC blocks. Next, each MAC variant is embedded in two representative DSP accelerators—a 6-tap FIR filter and an 8-point DCT/iDCT engine—to compare area, power, and accuracy across implementations. We omit a dedicated latency column at the accelerator level because each $4\times4$ MAC tiles in their implementations is performing one 256-cycle pass on a common 10ns clock, end-to-end latency is fixed and identical across all four MAC flavours, offering no additional discriminatory insight.

All designs, including bit-stream generation logic, were coded in Verilog RTL and synthesized with Synopsys Design Compiler with 32 nm EDK library from the Synopsys University Software Program under identical design constraints to extract area (µm²) and power (µW).  We report MAC-level accuracy using root-mean-square error (RMSE) and standard deviation of the error (SDE), and application-level quality via RMSE and peak signal-to-noise ratio (PSNR).





\subsection{Performance comparison for a MAC unit design}
\label{sec:area_power_accuracy_for_mac}
Tables~\ref{Table:Area_power_latency} (unipolar) and~\ref{Table:bipolar_Area_power_latency} (bipolar) compare the four $4 \times 4$ MAC variants in terms of area, power, latency, RMSE, and SDE. Each MAC unit computes the dot-product of two four-element vectors using 8-bit binary operands.
\begin{table}[ht!]\centering
\caption{Comparison for RMSE, SDE, area, power, and latency for a MAC unit - (unipolar 8-bit binary number)}\label{Table:Area_power_latency}
\scriptsize
\begin{tabularx}{\columnwidth}{X*{5}{>{\centering\arraybackslash}X}}
\toprule
\textbf {\mbox{MAC design}}  & \textbf{Area ($\mu m^2$)} 
& \textbf{Power ($\mu W$)} & \textbf{Latency ({\it ns})} & \textbf{RMSE ($\%$)} & \textbf{SDE  ($\%$)}\\
\midrule
CBSC MAC~\cite{Yu:DAC'21}  & 2202.67 & 63.91 & 2560  & 0.52 & 0.29\\
\midrule
HTC MAC~\cite{Maliha:ASPDAC’24}   & 609.68  & 23.84 & 2560  & 7.90  & 4.84\\ 
\midrule
\mbox{HTC MAC} \mbox{(EMBA)}  & 567.55  & 18.33 & 2560  & 0.52 & 0.35\\
\midrule
\mbox{HTC MAC} \mbox{(DTSA)}  & 570.46  & 22.50 & 2560  & 0.52 & 0.35\\
\bottomrule
\end{tabularx}
\end{table}

\begin{table}[ht!]\centering
\caption{Comparison for RMSE, SDE, area, power, and latency for a MAC unit - (bipolar 8-bit binary number)}\label{Table:bipolar_Area_power_latency}
\scriptsize
\begin{tabularx}{\columnwidth}{X*{5}{>{\centering\arraybackslash}X}}
\toprule
\textbf {\mbox{MAC design}}  & \textbf{Area ($\mu m^2$)} 
& \textbf{Power ($\mu W$)} & \textbf{Latency ({\it ns})} & \textbf{RMSE ($\%$)} & \textbf{SDE  ($\%$)}\\
\midrule
CBSC MAC~\cite{SimLee:DAC’17}  & 2553.77 & 86.79 & 2560  & 1.40 & 0.84\\
\midrule
HTC MAC~\cite{Maliha:ASPDAC’24}   & 839.78  & 35.78 & 2560  & 12.51  & 7.52\\ 
\midrule
\mbox{HTC MAC} \mbox{(EMBA)}  & 606.91  & 20.53 & 2560  & 2.09 & 1.40\\
\midrule
\mbox{HTC MAC} \mbox{(DTSA)}  & 830.78  & 27.62 & 2560  & 2.09 & 1.40\\
\bottomrule
\end{tabularx}
\end{table}
Table~\ref{Table:Area_power_latency} shows that EMBA and DTSA match the CBSC MAC’s RMSE of 0.52\% while reducing area by over 74\% and power by more than 64\%. Relative to the MUX-HTC MAC (RMSE 7.90\%), both EMBA and DTSA achieve around 94\% reduction in RMSE. Compared to MUX-HTC, EMBA reduces area and power by around 7\% and 23\%, respectively, while DTSA reduces area and power by 6.4\% and 5.6\%, respectively.

Table~\ref{Table:bipolar_Area_power_latency} shows that in the bipolar domain EMBA and DTSA achieve an RMSE of approximately 2.1\%, representing roughly an 83\% reduction compared to the MUX-HTC MAC (12.51\%).  While this approach CBSC’s superior RMSE of 1.40\%, CBSC still retains the accuracy lead.  In terms of hardware efficiency, EMBA reduces area and power by approximately 76\% each versus CBSC and by about 28\% and 43\%, respectively, versus MUX-HTC; DTSA reduces area and power by around 67\% and 68\% versus CBSC and by roughly 1\% and 23\% versus MUX-HTC.

The CBSC MAC attains superior accuracy in the bipolar domain by adapting its cycle count to the actual operand magnitude rather than using a fixed \(2^N\)-length bitstream.  Specifically, following~\cite{SimLee:DAC’17}, each $N$-bit two’s-complement operand is converted to sign–magnitude form, and an FSM generates a deterministic magnitude bitstream.  The sign of the product is obtained by XORing the operand signs, and an up–down counter accumulates +1/–1 over exactly $\bigl|\!2^{N-1}\cdot W\bigr|$ cycles- incrementing for ‘1’ bits and decrementing for ‘0’ bits. This on-demand, magnitude-driven counting skips unused cycles when $\bigl| W\bigr|$ is small and extends to capture larger products when $\bigl| W\bigr|$ is large, thereby minimizing bitstream variance and yielding the lower RMSE and higher PSNR.

\subsection{Application I: 6-Tap Finite Impulse Response Filter}
\label{sec:FIR}

To demonstrate the practical advantages of our newly proposed E-HTC MAC architectures, we implemented a 6-tap Gaussian-windowed FIR blur filter, chosen for its smooth low-pass response and minimal ringing artifacts~\cite{oppenheim2009dsp_ch7}. This FIR filter was applied to five images from the USC-SIPI image database~\cite{USC-SIPI}, which were selected to represent a range of textures and complexities. All filter coefficients were quantized as 8-bit unipolar fractional binary numbers to ensure compatibility with the bitstream computing framework. The quality of the filtered images was quantitatively evaluated using the root mean squared error (RMSE) and the peak signal-to-noise ratio (PSNR) in decibels (dB).


As detailed in Table~\ref{Table:FIR}, both EMBA and DTSA match CBSC’s accuracy. 
Across all five test images, both EMBA and DTSA designs consistently deliver a PSNR gain of 3--5\,dB and an RMSE reduction of approximately 30--45\% compared to the MUX-based design, underscoring the substantial accuracy benefits provided by our new deterministic addition methods.

In terms of hardware costs, the lower section of Table~\ref{Table:FIR} reports the area (in \(\mu\text{m}^2\)) and power consumption (in \(\mu\text{W}\)) for the FIR filter implementations. Relative to the MUX-based approach, EMBA reduces area by approximately 3\% and power by 13\%, while DTSA achieves 1\% area and 2\% power savings. Moreover, when compared to the CBSC MAC, both proposed E-HTC designs achieve an area reduction of more than 64\% along with power savings of
roughly 62\% for EMBA and around 57\% for DTSA. 

Fig.~\ref{fig:FIR_man} illustrates the filtering results for the ``Man'' image using all the mentioned approaches. Fig.~\ref{fig:man_original} shows the original image, while Fig.~\ref{fig:man_CBSC} displays the output from the CBSC, EMBA, and DTSA MACs, which all yield identical PSNR and RMSE and are therefore merged into a single image. Fig.~\ref{fig:man_MUX} corresponds to the existing MUX-based HTC MAC. As the figure reveals, the merged filtered output Image from CBSC/EMBA/DTSA is nearly indistinguishable from the original Image in terms of visual quality, confirming that our proposed designs achieve the same high accuracy as the state-of-the-art CBSC approach. In contrast, the MUX-based HTC MAC output exhibits noticeable differences, consistent with its higher RMSE and lower PSNR.
\begin{figure}[htbp!]
   \centering
   \begin{subfigure}{0.3\columnwidth}
       \includegraphics[width=\linewidth]{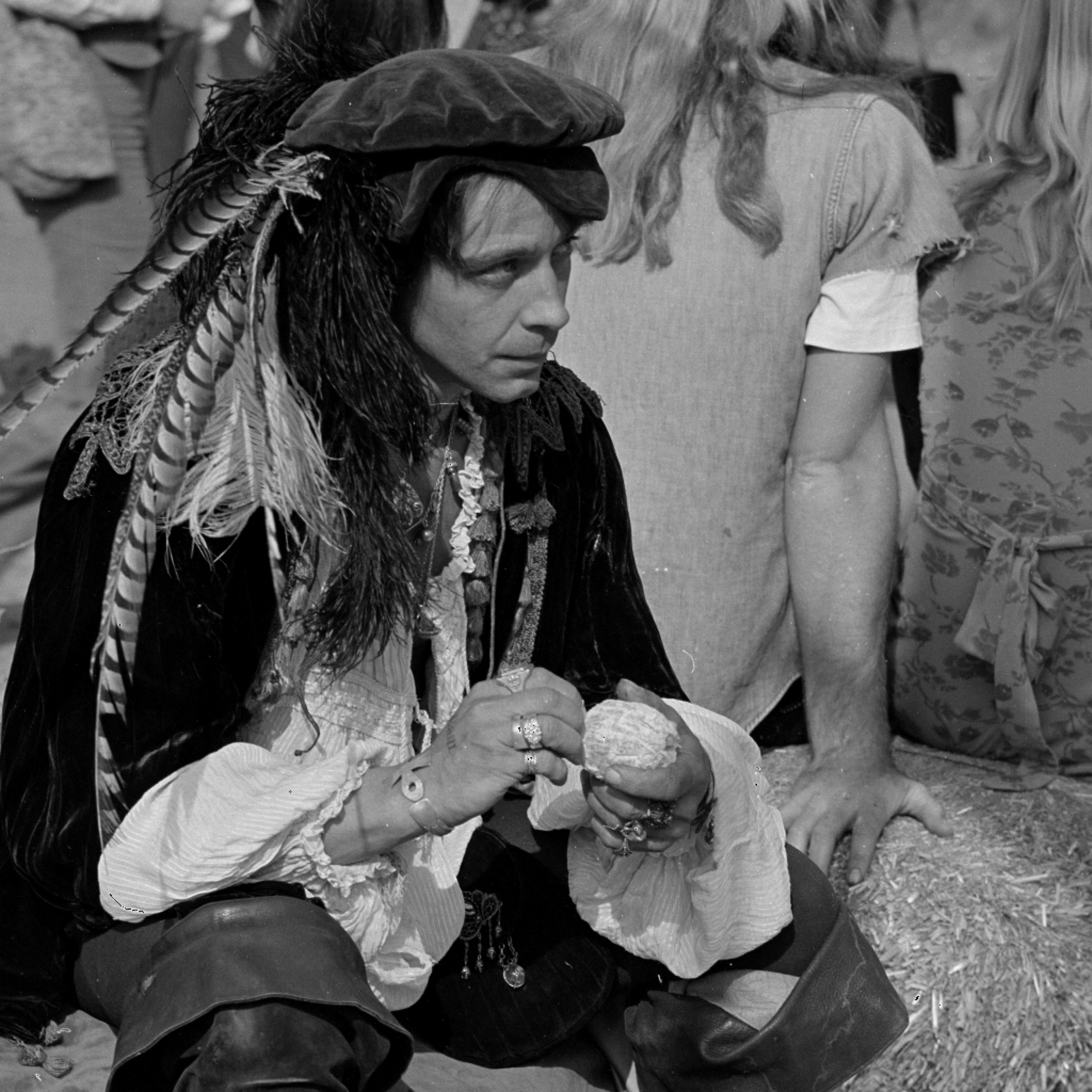}
       \caption{}
       \label{fig:man_original}
   \end{subfigure}
   \begin{subfigure}{0.3\columnwidth}
       \includegraphics[width=\linewidth]{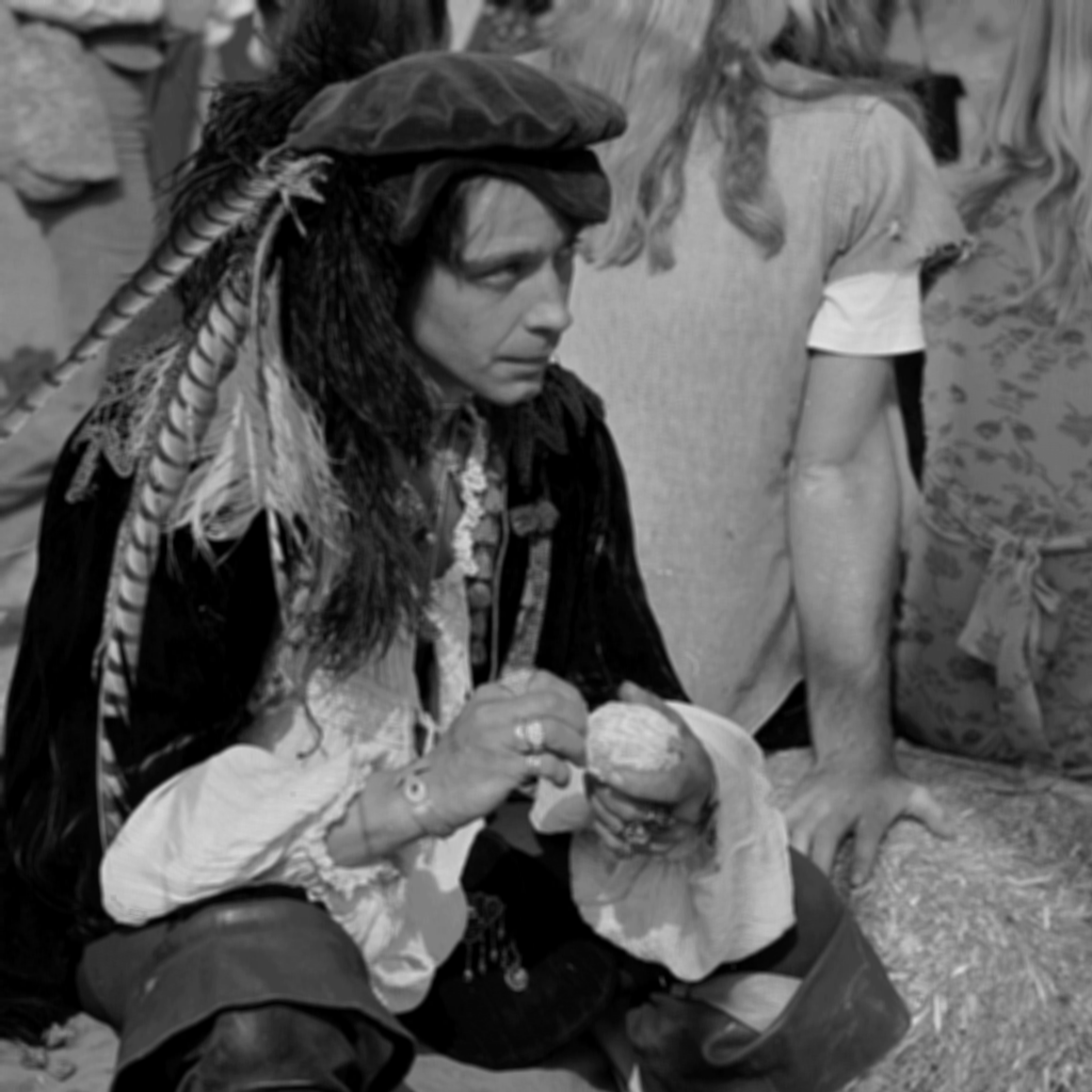}
       \caption{}
       \label{fig:man_CBSC}
   \end{subfigure}
   \begin{subfigure}{0.3\columnwidth}
       \includegraphics[width=\linewidth]{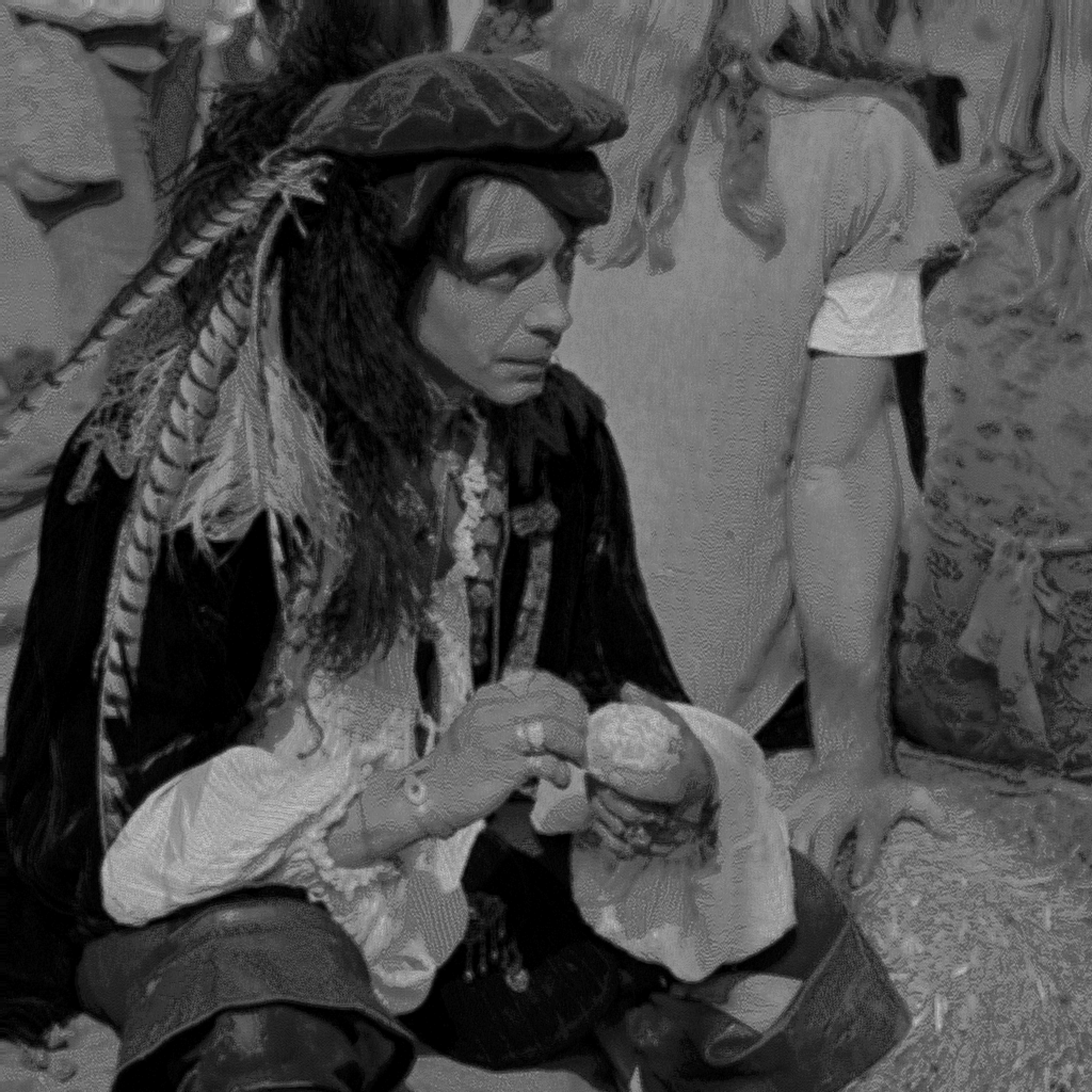}
       \caption{}
       \label{fig:man_MUX}
   \end{subfigure}

   

   \caption{The effect of applying Gaussian blurring filter on (a) Original Image, where the 6-tap unipolar filter is implemented with 
   (b) CBSC/EMBA/DTSA MAC (identical results);
   (c) HTC MAC (MUX-based)}
   \label{fig:FIR_man}
   \vspace{-0.1in}
\end{figure}


   

\begin{table}[ht!]
\centering
\caption{Performance comparison for 6-tap FIR-filter}
\label{Table:FIR}
\scriptsize  
\setlength{\tabcolsep}{1pt}  
\begin{tabularx}{\columnwidth}{%
  X                                      
  *{2}{>{\centering\arraybackslash}X}|   
  *{2}{>{\centering\arraybackslash}X}|   
  *{2}{>{\centering\arraybackslash}X}|   
  *{2}{>{\centering\arraybackslash}X}    
}
\toprule
\textbf{} 
& \multicolumn{2}{c|}{\textbf{CBSC MAC}}
& \multicolumn{2}{c|}{\shortstack{\textbf{HTC MAC}\\\textbf{(MUX based)}}}
& \multicolumn{2}{c|}{\shortstack{\textbf{HTC MAC}\\\textbf{(EMBA based)}}}
& \multicolumn{2}{c}{\shortstack{\textbf{HTC MAC}\\\textbf{(DTSA based)}}} \\
\cmidrule(lr){2-3}\cmidrule(lr){4-5}\cmidrule(lr){6-7}\cmidrule(lr){8-9}
\textbf{Image} 
& \textbf{PSNR (dB)} & \textbf{RMSE} 
& \textbf{PSNR (dB)} & \textbf{RMSE} 
& \textbf{PSNR (dB)} & \textbf{RMSE}
& \textbf{PSNR (dB)} & \textbf{RMSE}\\
\midrule
Boat   & 21.14 & 0.08 & 16.72 & 0.14 & 21.14 & 0.08 & 21.14 & 0.08 \\
Man    & 22.11 & 0.07 & 18.54 & 0.11 & 22.11 & 0.07 & 22.11 & 0.07 \\
Couple & 20.81 & 0.09 & 17.08 & 0.13 & 20.81 & 0.09 & 20.81 & 0.09 \\
Bridge & 19.62 & 0.10 & 16.61 & 0.14 & 19.62 & 0.10 & 19.62 & 0.10 \\
Clock  & 20.41 & 0.09 & 15.09 & 0.17 & 20.41 & 0.09 & 20.41 & 0.09 \\
\bottomrule
\bottomrule
\\
\textbf{Hardware cost}
& \textbf{Area ($\mu m^2$)} & \textbf{Power ($\mu W$)}
& \textbf{Area ($\mu m^2$)} & \textbf{Power ($\mu W$)}
& \textbf{Area ($\mu m^2$)} & \textbf{Power ($\mu W$)}
& \textbf{Area ($\mu m^2$)} & \textbf{Power ($\mu W$)} \\
\midrule
         & 3174.66 & 92.01
         & 1149.78 & 40.53
         & 1115.84 & 35.06
         & 1137.81 & 39.73\\
\bottomrule
\end{tabularx}
\vspace{-0.2in} 
\end{table}


   


\subsection{Application II: Discrete Cosine Transform}

We further evaluated our proposed E-HTC MAC designs for the bipolar approach in a widely used image compression tool, the Discrete Cosine Transform (DCT), where the coefficients can be negative. We implemented 8-point DCT accelerators with eight coefficients using our two new proposed E-HTC MAC architectures (EMBA and DTSA based) and compared with bipolar CBSC MAC~\cite{SimLee:DAC’17} and the existing MUX-based HTC MAC~\cite{Maliha:ASPDAC’24}. This example compares the performance of the two new proposed E-HTC MAC designs in bipolar encoding designs. 


   


\begin{figure}[htbp]
  \centering
  \makebox[\columnwidth][c]{%
    \begin{subfigure}{0.3\columnwidth}
      \includegraphics[width=\linewidth]{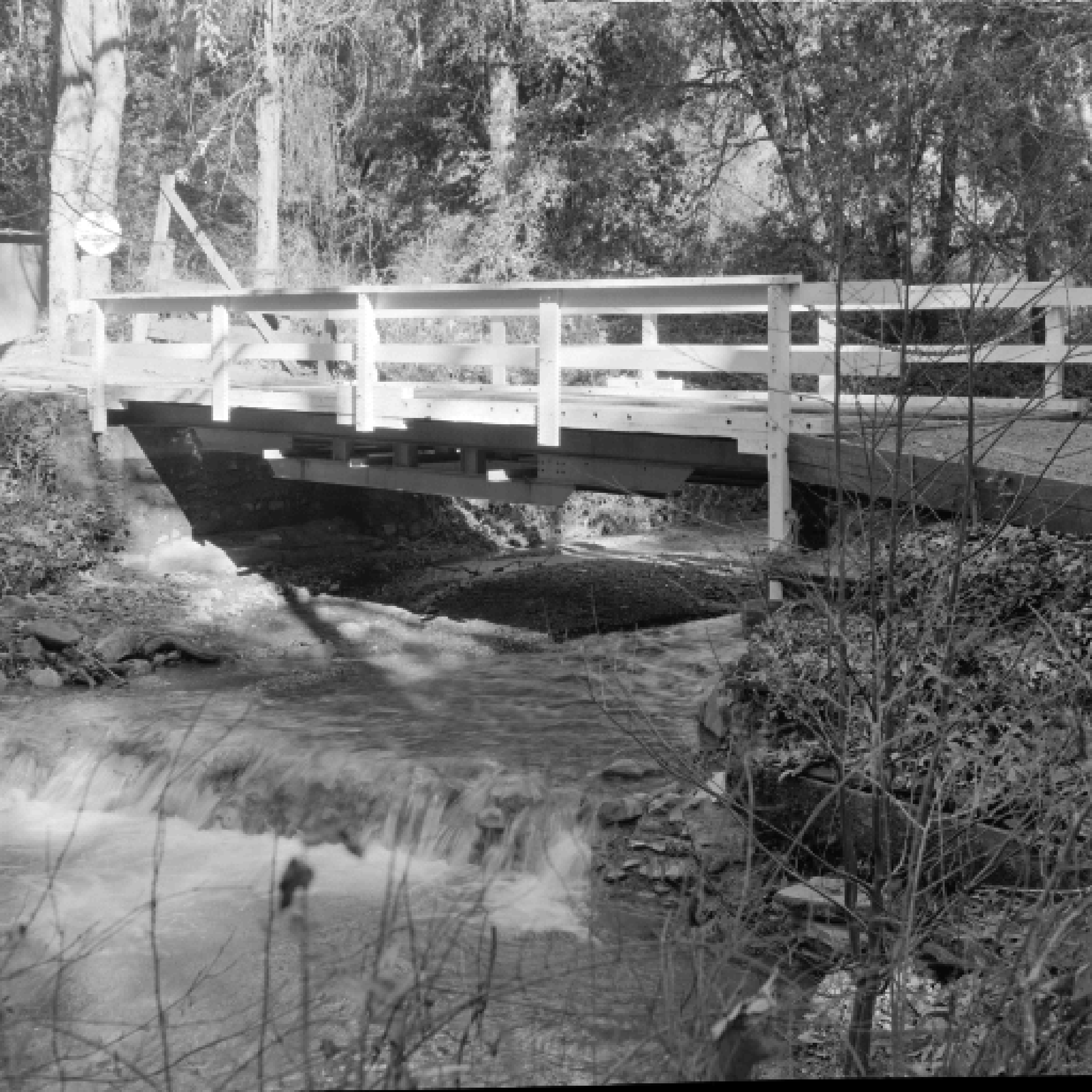}
      \caption{}
      \label{fig:bridge_original}
    \end{subfigure}
    \hspace{0.05\columnwidth}
    \begin{subfigure}{0.3\columnwidth}
      \includegraphics[width=\linewidth]{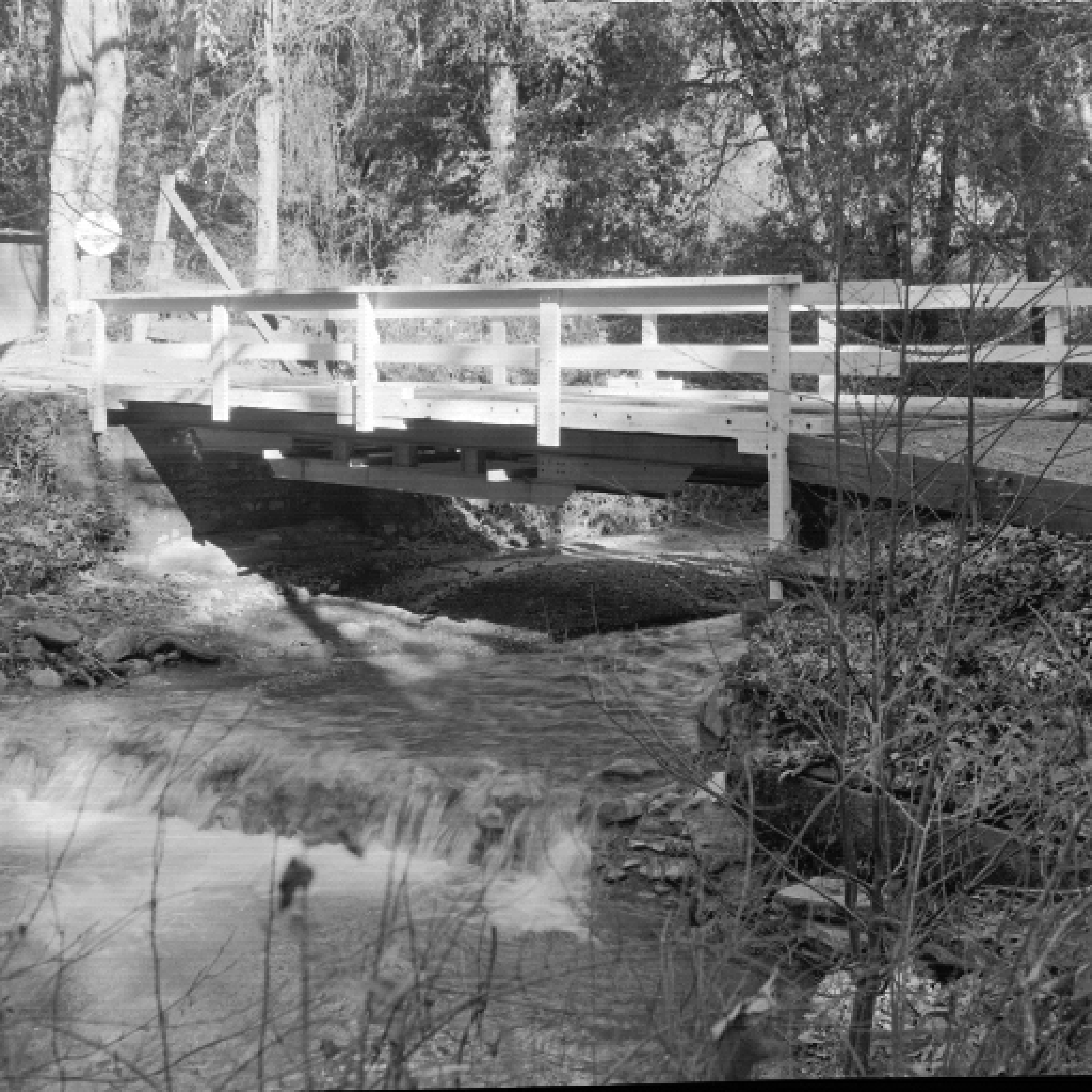}
      \caption{}
      \label{fig:bridge_CBSC}
    \end{subfigure}
  }

  \vspace{0.2em}

  \makebox[\columnwidth][c]{%
    \begin{subfigure}{0.3\columnwidth}
      \includegraphics[width=\linewidth]{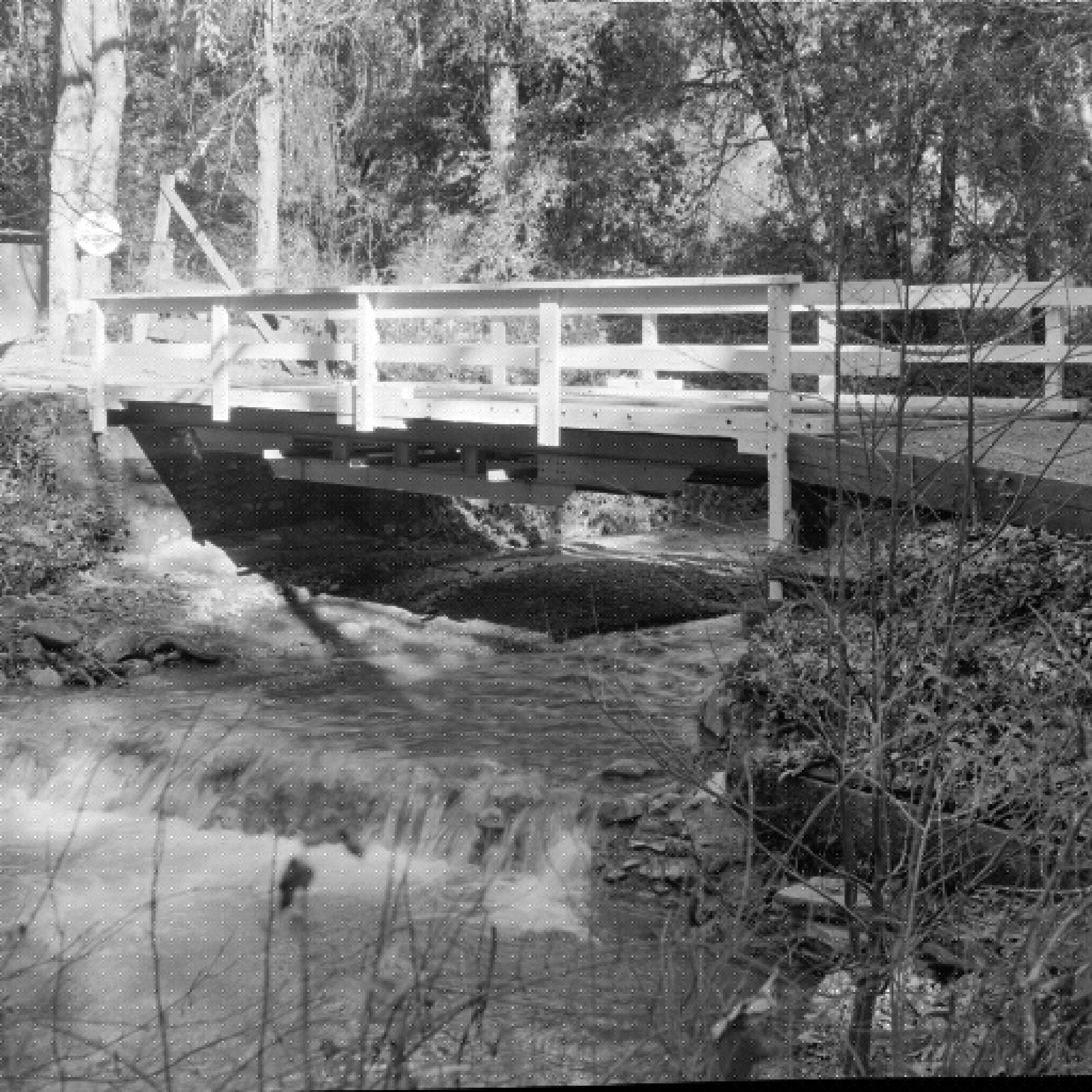}
      \caption{}
      \label{fig:bridge_EMBA}
    \end{subfigure}
    \hspace{0.05\columnwidth}
    \begin{subfigure}{0.3\columnwidth}
      \includegraphics[width=\linewidth]{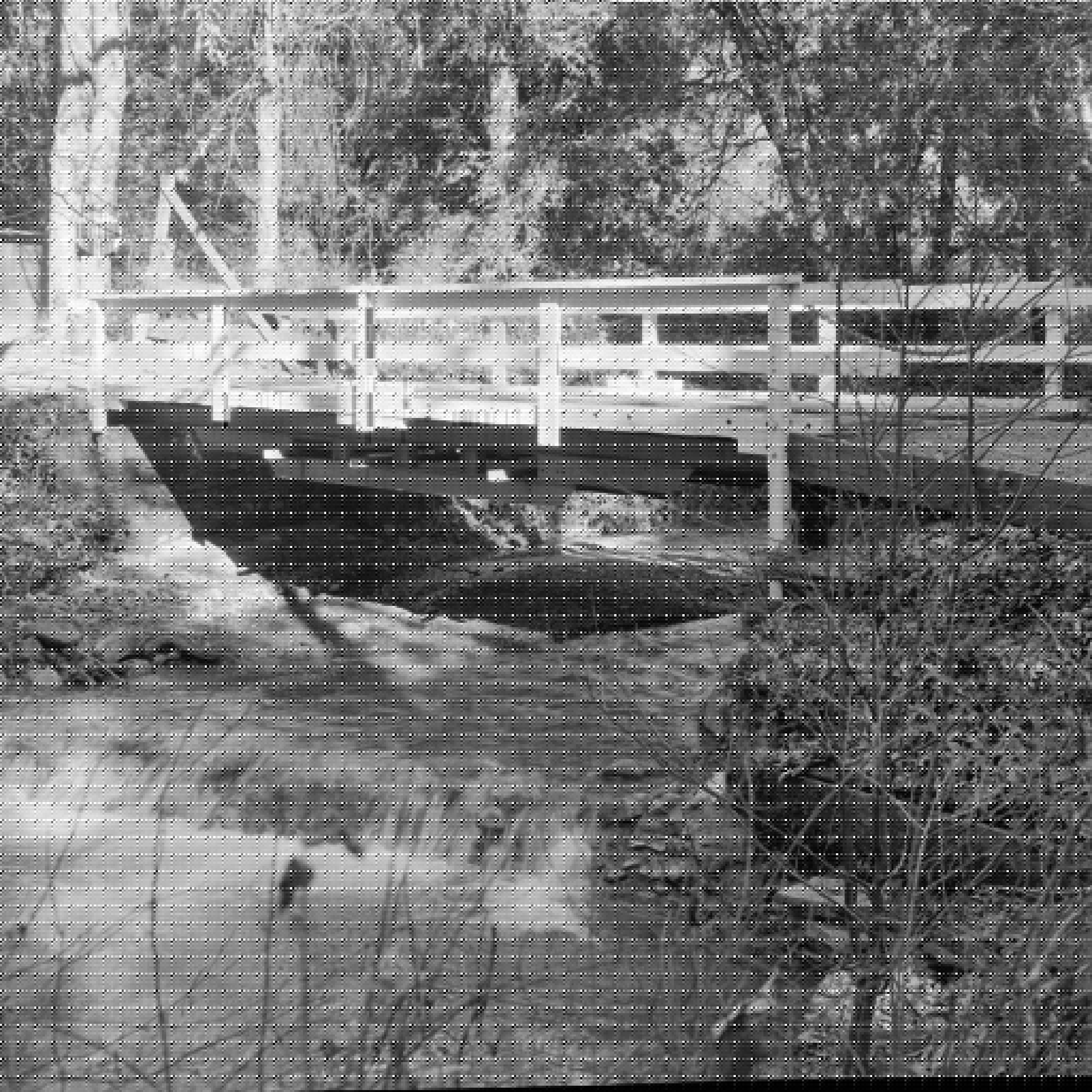}
      \caption{}
      \label{fig:bridge_MUX}
    \end{subfigure}
  }

  \caption{The discrete cosine filter applied to (a) original image of a bridge, where filter MAC is implemented with 
   (b) CBSC MAC,
   (c) HTC MAC (EMBA/DTSA; identical results),
   (d) HTC MAC (MUX based).}
  \label{fig:DCT_bridge}
  \vspace{-0.1in}
\end{figure}

We first transformed five images of varying resolutions from the USC-SIPI-miscellaneous dataset~\cite{USC-SIPI}. All images and the 8-point DCT-II coefficients were quantized to 8-bit signed fractional binary numbers before applying the filter accelerator to the images. The transformed images were then filtered back to the original domain using an 8-point inverse Discrete Cosine Transform (IDCT) filter. 
The quality of the transformed images was evaluated by computing the peak signal-to-noise ratio (PSNR) in dB and the root mean square error (RMSE) of the image transformed back to the original form by the IDCT filter.

Table \ref{Table:DCT} presents the error metrics and hardware implementation costs of these designs. The CBSC MAC was implemented using a bipolar CBSC multiplier architecture to multiply two signed numbers, as described in \cite{SimLee:DAC’17}, with exact binary adders summing the multiplication results. The 8-point DCT filter was then built from two 4×1 CBSC MACs, each performing four CBSC multiplications and three binary additions.

The proposed new E-HTC MAC-based DCT filter accelerator was implemented using the same MAC architectures discussed in Section~\ref{proposed_idea_sec}. The idea of generating a regulated and temporal bit stream for the bipolar remains the same as described in~\cite{Maliha:ASPDAC’24} and also reviewed in Section~\ref{sec:review_of_htc}. Our proposed E-HTC MAC architectures incorporate four bipolar HTC multipliers and one adder unit (EMBA/DTSA) to sum up the results of the four multiplications. Two four-input HTC MACs are utilized to implement the 8-point approximate DCT filter.

As shown in Table~\ref{Table:DCT}, both our newly proposed E-HTC MAC designs yield substantially higher PSNR values than the existing MUX-based HTC design while consuming considerably less area and power. 
Across all five test images, EMBA and DTSA consistently show improvements of about 10--12\,dB in PSNR and roughly 70--75\% reductions in RMSE relative to the MUX-based design, confirming the significant accuracy boost provided by our new deterministic summation methods.


Although, the CBSC MAC-based DCT filter achieves
higher PSNR than our proposed E-HTC designs because of the reason described in Section~\ref{sec:area_power_accuracy_for_mac}. However, this accuracy comes at a hardware cost as EMBA and DTSA offer more than 10\% reduction in the area and draw power of 0.55--0.60\,mW, which is nearly 90\% power savings than CBSC (5.58\,mW). These results demonstrate that E-HTC MAC designs provide the best trade‑off between accuracy and hardware efficiency for DCT-based image compression, substantially surpassing the MUX-based baseline while retaining far lower power and area costs than CBSC.

Fig.~\ref{fig:DCT_bridge} illustrates the results for the ``Bridge'' image using all the mentioned approaches. Fig.~\ref{fig:bridge_original} shows the original image, while Fig.~\ref{fig:bridge_CBSC} displays the output from the CBSC MAC, Fig.~\ref{fig:bridge_EMBA} corresponds to our EMBA/DTSA based MAC and Fig.~\ref{fig:bridge_MUX} correspond to the existing MUX-based HTC MAC. As the figure reveals, our proposed E-HTC design-based DCT filter retains the quality of the original image with a PSNR range of approximately 30\,dB, which is generally sufficient for such operations.
 
\begin{table}[ht!]
\centering
\caption{Performance comparison for 8-point DCT}
\label{Table:DCT}
\scriptsize
\setlength{\tabcolsep}{1pt}
\begin{tabularx}{\columnwidth}{%
  X
  *{2}{>{\centering\arraybackslash}X}|
  *{2}{>{\centering\arraybackslash}X}|
  *{2}{>{\centering\arraybackslash}X}|
  *{2}{>{\centering\arraybackslash}X}
}
\toprule
\textbf{}
& \multicolumn{2}{c|}{\textbf{CBSC MAC}}
& \multicolumn{2}{c|}{\shortstack{\textbf{HTC MAC}\\\textbf{(MUX-based)}}}
& \multicolumn{2}{c|}{\shortstack{\textbf{HTC MAC}\\\textbf{(EMBA-based)}}}
& \multicolumn{2}{c}{\shortstack{\textbf{HTC MAC}\\\textbf{(DTSA-based)}}} \\
\cmidrule(lr){2-3}\cmidrule(lr){4-5}\cmidrule(lr){6-7}\cmidrule(lr){8-9}
\textbf{Image}
& \textbf{PSNR (dB)} & \textbf{RMSE}
& \textbf{PSNR (dB)} & \textbf{RMSE}
& \textbf{PSNR (dB)} & \textbf{RMSE}
& \textbf{PSNR (dB)} & \textbf{RMSE}\\
\midrule
Boat   & 40.37 & 2.44 & 18.69 & 29.63 & 30.88 & 7.28 & 30.88 & 7.28 \\
Man    & 39.16 & 2.81 & 18.10 & 31.74 & 30.62 & 7.51 & 30.62 & 7.51 \\
Couple & 40.89 & 2.30 & 18.52 & 30.25 & 30.48 & 7.63 & 30.48 & 7.63 \\
Bridge & 39.98 & 2.55 & 18.49 & 30.33 & 30.42 & 7.68 & 30.42 & 7.68 \\
Clock  & 41.01 & 2.26 & 20.18 & 24.99 & 31.62 & 6.69 & 31.62 & 6.69 \\
\bottomrule
\bottomrule
\\
\textbf{Hardware cost}
& \textbf{Area ($\mu m^2$)} & \textbf{Power ($\mu W$)}
& \textbf{Area ($\mu m^2$)} & \textbf{Power ($\mu W$)}
& \textbf{Area ($\mu m^2$)} & \textbf{Power ($\mu W$)}
& \textbf{Area ($\mu m^2$)} & \textbf{Power ($\mu W$)} \\
\midrule
         & 35204.10 & 5580
         & 34626.52 & 646.18
         & 30570.96 & 553.48
         & 31716.99 & 596.06\\
\bottomrule
\end{tabularx}
\vspace{-0.1in} 
\end{table}


   


\section{Conclusion}
\label{sec:conclusion}
In this work, we introduce an accuracy-enhanced Hybrid Temporal Computing (E-HTC) framework that replaces the stochastic, MUX-based adder of existing HTC designs with two new
deterministic summation schemes: the Exact Multiple-input Binary Accumulator (EMBA), which directly accumulates multiple bitstream inputs in binary form and the Deterministic Threshold-based Scaled Adder (DTSA), which utilizes threshold-based logic to perform scaled addition. We presented the detailed architecture of the proposed adders and then developed corresponding $4\times4$ MAC arithmetic units and demonstrated them on two applications: Finite Impulse Response (FIR) filter and DCT/iDCT engine design. Experiment results showed that for a $4\times4$ MAC,
EMBA and DTSA match the 0.52\% RMSE of state-of-the-art
CBSC designs in unipolar coding yet reduce power by over 64\% and area by over
74\%. Relative to existing MUX-based HTC MAC, it is 94\% RMSE improvement along with power and area benefits
of around 23\% and 7\%, respectively, by EMBA and 5.6\% and 6.4\% by DTSA. In bipolar mode, the E-HTC MAC achieves an RMSE of 2.09\%---an 83\% improvement over the bipolar MUX-based HTC MAC and approaching the bipolar CBSC MAC’s RMSE of  1.40\%---while cutting area and power by roughly up to 28\% and 43\% versus MUX-HTC and by  up to about 76\% each versus CBSC. For the FIR filter design, both schemes
provide 3--5\,dB PSNR gain (35--45\% RMSE reduction) over the MUX-based design with similar accuracy to CBSC while
saving up to 13\% power and 3\% area over the MUX-based design. Compared to CBSC, the proposed design yields area
and power reductions of nearly 65\% and 62\%, respectively. In the DCT/iDCT application, EMBA and DTSA deliver 10–13\,dB PSNR improvements (70--75\% RMSE drop) over MUX-
based HTC, along with roughly 90\% power and more than 10\% area improvements over CBSC.


\bibliographystyle{ieeetr}
\bibliography{./vsclab_bib_database/security,./vsclab_bib_database/emergingtech,./vsclab_bib_database/thermal_power,./vsclab_bib_database/mscad_pub,./vsclab_bib_database/interconnect,./vsclab_bib_database/stochastic,./vsclab_bib_database/simulation,./vsclab_bib_database/modeling,./vsclab_bib_database/reduction,./vsclab_bib_database/misc,./vsclab_bib_database/architecture,./vsclab_bib_database/reliability,./vsclab_bib_database/thermal,./vsclab_bib_database/neural_network,./vsclab_bib_database/machine_learning,./vsclab_bib_database/realtime,./vsclab_bib_database/approximate_comp}
\end{document}